\documentclass[onecollarge,final]{svjour2}

\smartqed

\usepackage [dvips]{graphicx}
\usepackage{amssymb}
\usepackage{epsfig}
\usepackage{graphicx}
\usepackage{verbatim}

\usepackage[T1]{fontenc}

\usepackage{amsmath}
\usepackage{enumerate}
\usepackage{graphicx}
\usepackage{mathbbol}
\usepackage{amsfonts}
\newcommand{\gras}[1]{\boldsymbol{#1}}

\begin{document}

%\preprint{}

\title{Extremes of $N$ vicious walkers for large $N$: application to the directed polymer and KPZ interfaces}% repeat the \author .. \affiliation  etc. as needed


\author{Gr\'egory Schehr}

\institute{G. Schehr \at Laboratoire de Physique Th\'eorique, Universit\'e de
  Paris-Sud,  91405 Orsay France \\  Laboratoire de Physique Th\'eorique et Mod\`eles
  Statistiques, Universit\'e Paris-Sud, B\^at. 100, 91405 Orsay,
France \\
\email{schehr@u-psud.fr}
}

\date{\today}

\maketitle

\begin{abstract}
We compute the joint probability density function (jpdf) $P_N(M, \tau_M)$ of the maximum $M$ and its position $\tau_M$ for $N$ non-intersecting Brownian excursions, on the unit time interval, in the large $N$ limit. For $N \to \infty$, this jpdf is peaked around $M = \sqrt{2N}$ and $\tau_M = 1/2$, while the typical fluctuations behave for large $N$ like $M - \sqrt{2N} \propto s N^{-1/6}$ and $\tau_M - 1/2 \propto w N^{-1/3}$ where $s$ and $w$ are correlated random variables. One obtains an explicit expression of the limiting jpdf $P(s,w)$ in terms of the Tracy-Widom distribution for the Gaussian Orthogonal Ensemble (GOE) of Random Matrix Theory and a psi-function for the Hastings-McLeod solution to the Painlev\'e II equation. Our result yields, up to a rescaling of the random variables $s$ and $w$, an expression for the jpdf of the maximum and its position for the Airy$_2$ process minus a parabola. This latter describes the fluctuations in many different physical systems belonging to the Kardar-Parisi-Zhang (KPZ) universality class in $1+1$ dimensions. In particular, the marginal probability density function (pdf) $P(w)$ yields, up to a model dependent length scale, the distribution of the endpoint of the directed polymer in a random medium with one free end, at zero temperature. In the large $w$ limit one shows the asymptotic behavior $\log P(w) \sim -  w^3/12$.

\end{abstract}

\section{Introduction}

Despite several decades of research, there exist very few exact results for finite dimensional disordered systems, beyond mean-field or phenomenological arguments. One important model is the directed polymer in a random medium (DPRM) which has been an active area of research in statistical physics for the past three decades~\cite{halpin_review}. Apart from the fact that it is a simple toy model of disordered systems, this problem has important links to a wide variety of other problems in physics, such as interface fluctuations and pinning~\cite{huse_henley,kardar_dprm}, growing interface model of the Kardar-Parisi-Zhang (KPZ) variety~\cite{kpz}, Burger's turbulence~\cite{burgers}, spin glasses~\cite{mezard_dprm} or in biological sequence matching problems~\cite{hwa_lassig}. The DPRM is also a minimal model, yet non trivial, of great interest in the context of disordered elastic systems which have found many experimental realizations ranging from domain walls in random ferromagnets \cite{lemerle} to wetting on rough substrates~\cite{moulinet} etc.

\begin{figure}
\centering
\includegraphics[width=0.5\linewidth]{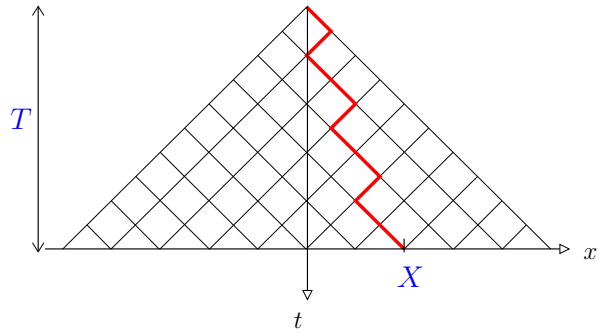}
\caption{Directed polymer of length $T$ on a square lattice. On each bond, there is a random variable $\epsilon_{ij}$.}\label{fig_polymer}
\end{figure}

For concreteness we consider a directed polymer on a square lattice, as depicted in Fig.~\ref{fig_polymer}. On each site with coordinate $(i,j)$ there is a random energy $\epsilon_{ij}$ drawn independently from site to site, which is a quenched random variable. Here we consider the ensemble of directed walks of length $T$ (as depicted in Fig. \ref{fig_polymer}) starting from the origin, making step diagonaly left or right, and ending at any point on the line depicted on Fig. \ref{fig_polymer} as the $x$-axis: this is the DPRM with one free end, or in point to line geometry. The total energy $E({\cal W})$ for any directed walk ${\cal W}$ is just the sum of site energies along the walk $E({\cal W})= \sum_{(i,j) \in {\cal W}} \epsilon_{ij}$ and one is interested in the path having minimum energy, that is the ground state or the optimal path. There are two natural quantities describing this optimal path: its energy, $E_{\rm opt}(T)$ and the position of the endpoint $X(T)$~\cite{huse_henley,kardar_dprm} (see Fig. \ref{fig_polymer}). Earlier studies of this kind of model focused on the roughness of this ground state configuration, which is a geometrical characteristic, and it was found that~\cite{huse_henley,kardar_dprm}
\begin{eqnarray}\label{roughness}
\overline {X_{}^2}(T) \propto T^{4/3} \;, \; T \gg 1 \;,
\end{eqnarray}
where $\overline{...}$ stands for an average over the disorder, that is over the random variables $\epsilon_{ij}$. 

Similarly the two first moments of the energy of the ground state $E_{\rm opt}(T)$ were found to scale~\cite{huse_henley,kardar_dprm} like
\begin{eqnarray}\label{energy}
\overline{E_{\rm opt}}(T) \sim a \, T \;, \; \overline{E_{\rm opt}^2}(T) - \overline{E_{\rm opt}}^2(T)\propto T^{2/3} \;,\; T \gg 1 \;.
\end{eqnarray}
These behaviors (\ref{roughness}, \ref{energy}) were later proved rigorously for one specific model of (discrete) directed polymer~\cite{johansson_dprm}. Quite remarkably, more recent studies were able to obtain the  full distribution of the energy $E_{\rm opt}(T)$ which takes the form, for $T \gg 1$
\begin{eqnarray}\label{scaling_eopt}
E_{\rm opt}(T) = a T + b \chi T^{1/3} \;, 
\end{eqnarray}   
where $a, b$ are non-universal constants, depending on the precise model of DPRM under consideration, while $\chi$ is a random variable such that
${\rm Pr}(\chi \leq s) = {\cal F}_1(s)$ where ${\cal F}_1(s)$ is the Tracy-Widom distribution which describes the typical fluctuations of the largest eigenvalue of the Gaussian Orthogonal Ensemble (GOE) \cite{TW96}. This was shown, rather indirectly, in Ref. \cite{johansson_dprm} using a relation between certain observables of the DPRM in point to point geometry and associated ones in point to line geometry~\cite{krug_pra} for which the limiting distribution had been evaluated~\cite{baik_rains}. In the context of stochastic growth models, namely for the Polynuclear Growth Model in flat geometry, this was shown in Ref. \cite{spohn_praehofer_prl}. This was then also shown, later, by a direct computation using a relation with non-intersecting Brownian motions~\cite{forrester_npb,liechty} and more recently in Ref. \cite{quastel_jpdf} using probabilistic tools. Note that in the case where the end of the polymer is fixed (in point to point geometry), the same scaling (\ref{scaling_eopt}) holds, with different constants $a', b'$, and the distribution of $\chi$ is also different. It was indeed shown~\cite{johansson_dprm,spohn_praehofer_prl,johansson_2000,spohn_praehofer} that in this case the random variable $\chi$ is distributed according to ${\cal F}_2$, the Tracy-Widom distribution corresponding to the Gaussian Unitary Ensemble (GUE)~\cite{TW94a}, establishing also a connection with the seminal paper of Baik, Deift and Johansson \cite{BDJ} on the longest increasing subsequence of a random permutation~\cite{satya_review}. Recent approaches have also studied continuum models of directed polymers both in point to point \cite{calabrese_epl,dotsenko_epl,dotsenko_jstat,sasamoto_spohn1,sasamoto_spohn2,sasamoto_spohn3,amir_cmp,prolhac_spohn1,prolhac_spohn2} and in point to line geometries~\cite{calabrese_flat,calabrese_flat_long}.  

For the directed polymer in point to line geometry, {\it i.e.} with one free end as in Fig. \ref{fig_polymer}, it was further shown~\cite{johansson_dprm} that the fluctuations of $E_{\rm opt} - \overline{E_{\rm opt}}$ and $X$ are described, up to a non-universal rescaling, 
by the statistical properties of the following process $Y(u)$
\begin{eqnarray}\label{def_y}
Y(u) = {\cal A}_2(u) - u^2 \;,
\end{eqnarray}
where ${\cal A}_2(u)$ is the Airy$_2$ process \cite{spohn_praehofer}. Indeed, one has
\begin{eqnarray}\label{long}
\lim_{T \to \infty}\frac{T^{-\frac{1}{3}}}{e_0}(E_{\rm opt}(T) - \overline{E_{\rm opt}}(T)) = \max_{u \in \mathbb{R}} Y(u) \equiv m\;,
\end{eqnarray}
while
\begin{eqnarray}\label{trans}
\lim_{T \to \infty} \frac{T^{-\frac{2}{3}}}{\xi} X = \arg \max_{u \in \mathbb{R}} Y(u) \equiv t\;,
\end{eqnarray}
which is simply the position at which this process $Y(u)$ (\ref{def_y}) reaches its maximum. In Eqs. (\ref{long}, \ref{trans}), the constant $e_0$ and $\xi$ are non-universal amplitudes, which depend on the microscopic details of the model under consideration. In other words, these relations (\ref{long}, \ref{trans}) mean that the joint probability density function (jpdf) $\hat P(m,t)$ of the rescaled energy and the rescaled position of the endpoint are given by the jpdf of the maximum of the Airy$_2$ process minus a parabola (\ref{def_y}) and its  position. In a recent paper~\cite{quastel_jpdf}, Moreno Flores, Quastel and Remenik have computed this jpdf using rigorous probabilistic tools concerning determinantal point processes. They obtained an 
expression [see Eq. (\ref{expr_quastel}) below] in terms of the product of ${\cal F}_1$ and a double integral involving in particular the integral operator ${\cal B}_m$ [see Eq. (\ref{kernel_ai1}) below], which also enters the expression of ${\cal F}_1$ (\ref{f1_fredholm}) as a Fredholm determinant~\cite{ferrari_spohn}.

\begin{figure}
\centering
\includegraphics[width=0.6\linewidth]{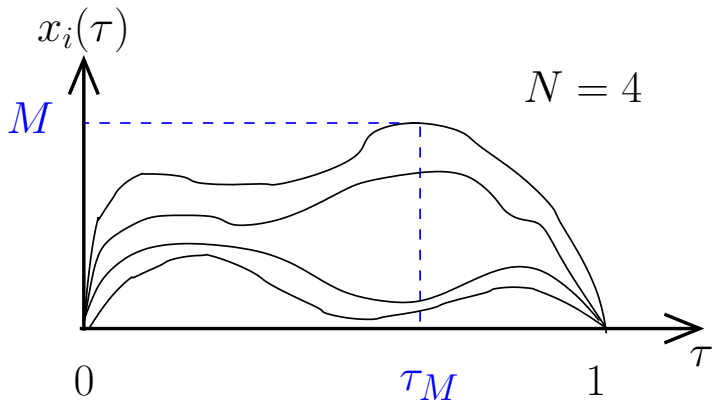}
\caption{Cartoon of $N$ non-intersecting Brownian excursions on the unit time interval. Here we study the jpdf $P_N(M,\tau_M)$ in the large $N$ limit.}\label{fig_excursions}
\end{figure}
Here we follow a completely different approach and use the fact that $Y(u)$ also describes the fluctuations of the top path of $N$ non-intersecting Brownian excursions in the large $N$ limit \cite{spohn_praehofer,tracy_excursions,corwin_hammond}. We remind that a Brownian excursion is a Brownian motion starting and ending at the same point, and conditioned to stay positive in-between. We obtain an explicit expression for the jpdf, different from the one obtained in Ref. \cite{quastel_jpdf} as the product of ${\cal F}_1$ and an integral involving a function which can be expressed in terms of the so-called psi-function \cite{FN_80,BI03} for the Hastings-McLeod solution of the Painlev\'e II equation [see Eq. (\ref{explicit_h1_intro}) below]. Following previous works \cite{forrester_npb,schehr_prl,rambeau_epl,rambeau_pre}, we thus consider the problem of $N$ non-colliding Brownian motions conditioned to stay positive on the unit time interval~(see Fig. \ref{fig_excursions})
\begin{eqnarray}
0 \leq x_1(\tau)<\cdots < x_N(\tau) \;, \; \forall \; \tau \in [0,1] \;.
\end{eqnarray}
We focus on special configurations called "watermelons with a wall" where  
the walkers start, at time $\tau =0$ and end, at time  $\tau =1$, at the origin~(see Refs. \cite{tracy_excursions,katori_jstatphys} for a rigorous definition of this model). We focus on the maximal height $M$ and the time at which this maximum is reached (see Fig. \ref{fig_excursions}):   
\begin{eqnarray}\label{def_Mtau}
M = \max_{0 \leq \tau \leq 1} x_N(\tau) \;, \; x_N(\tau_M) = M \;.
\end{eqnarray}
The distribution of $M$ was computed for $N=2$ in Ref. \cite{katori_jstatphys} and then for any value of $N$ in Ref. \cite{schehr_prl,kobayashi_pre,feierl}. Remarkably, it was shown \cite{forrester_npb} that the cumulative distribution of $M$, $F_N(M)$, coincides, up to a prefactor, with the partition function of Yang-Mills theory on the sphere, with the gauge group ${\rm Sp}(2N)$ \cite{GM94,CNS96} (see also Ref.~\cite{deharo} for the relation between non-intersecting Brownian motions and gauge field theories). Hence $F_N(M)$ exhibits a third order phase transition \cite{GW80,W80,DK93} for $M = \sqrt{2N}$, separating the left tail of the distribution (which corresponds to the regime of strong coupling in the Yang-Mills theory) from the right tail (which corresponds to the regime of weak coupling in the Yang-Mills theory). In the critical regime, for $M$ close to $\sqrt{2N}$, which is described by a double scaling limit~\cite{GM94,PS90a}, it was shown that $F_N(M)$, appropriately shifted and scaled, converges to ${\cal F}_1$~\cite{forrester_npb} [see also Eq. (\ref{asympt_FN}) below]. A rigorous proof of this convergence was recently given in Ref. \cite{liechty} using Riemann Hilbert techniques. 

The jpdf of $M$ and $\tau_M$ (\ref{def_Mtau}), $P_N(M, \tau_M)$ in Eq. (\ref{start_expr_first}), which is the starting point of our analysis, was computed in Ref. \cite{rambeau_epl,rambeau_pre}. In this paper we provide a large $N$ analysis of this jpdf. Our main results can be summarized as follows. We show that
\begin{eqnarray}\label{result_intro}
\lim_{N \to \infty} 2^{-\frac{9}{2}}N^{-\frac{1}{2}} P_N(\sqrt{2N} + 2^{-\frac{11}{6}}\, s \,N^{-\frac{1}{6}}, \frac{1}{2} + 2^{-\frac{8}{3}} \, w \, N^{-\frac{1}{3}} ) = {P}(s,w) \;, 
\end{eqnarray}
where the jpdf ${P}(s,w)$ is given by~\footnote{The factor $2^{-9/2}$ in the formula (\ref{result_intro}) ensures the normalization of the jpdf $P(s,w)$.}
\begin{eqnarray}\label{explicit_P}
{P}(s,w) = \frac{4}{\pi^2} {\cal F}_1(s) \int_{s}^\infty h(x,w) h(x,-w) \, dx \;,
\end{eqnarray}
where ${\cal F}_1(s)$ is the Tracy-Widom distribution for $\beta=1$ [see Eq. (\ref{F1_intro}) below] and $h(s,w)$ is given by
\begin{eqnarray}\label{explicit_h1_intro}
h(s,w) =  \int_{0}^\infty  \zeta \Phi_2(\zeta,s) e^{- w \zeta^2} \, d\zeta  \;, 
\end{eqnarray}
in terms of $\Phi_2(\zeta,s)$ which is one of the two components of the psi-function for the Hastings-McLeod solution to the Painlev\'e II equation~\cite{FN_80,BI03}. To define this psi-function one considers a Lax pair associated to the Hastings-McLeod solution of the Painlev\'e II equation \cite{BI03}, {\it i.e.} the following system of linear differential equations for a two-dimensional vector $\Psi \equiv \Psi(\zeta,s)$,
 \begin{eqnarray}\label{system_lax_intro}
 \frac{\partial}{\partial \zeta} \Psi = A \Psi \;, \; \frac{\partial}{\partial s} \Psi = B \Psi \;,
 \end{eqnarray} 
 where the $2 \times 2$ matrices $A \equiv A(\zeta, s)$ and $B \equiv B(\zeta,s)$ are given by
 \begin{eqnarray}\label{matrice_AB}
 A(\zeta,s) = \left( 
 \begin{array}{c c}
 4 \zeta q &  4 \zeta^2 + s + 2q^2 + 2r\\
 -4 \zeta^2 - s - 2 q^2 + 2r & -4 \zeta q
 \end{array}\right) \;, \;
  B(\zeta,s) = \left( 
 \begin{array}{c c}
 q &  \zeta \\
 -\zeta & - q
 \end{array}\right) \;,
 \end{eqnarray}
 with $q \equiv q(s)$ and $r \equiv r(s)$, where $q(s)$ is the Hastings-McLeod solution of the Painlev\'{e} 
\textrm{II} differential equation
\begin{equation}
q''(s) = 2q^3(s) + s q(s) \;, \; q(s)\mathop{\sim}\limits_{s\rightarrow \infty} {\rm Ai}(s) \;,
\label{qdiff}
\end{equation}
where ${\rm Ai}(s)$ is the Airy function~\cite{abramowitz}. These matrices $A$ and $B$ constitute a Lax pair associated to $q(s)$: this means that the compatibility equation of this system (\ref{system_lax_intro}), {\it i.e.} $\partial_s \partial_\zeta \Psi = \partial_\zeta \partial_s \Psi$, 
 is that $q(s)$ satisfies the Painlev\'e-II equation (\ref{qdiff}) and that $r(s) = q'(s)$. The function $\Phi_2(\zeta,s)$ which enters into the expression of the jpdf ${P}(s, w)$ (\ref{explicit_P},\ref{explicit_h1_intro}) denotes one component of the unique solution $\left(\Phi_1(\zeta,s), \Phi_2(\zeta,s)\right)$
of the Lax pair (\ref{system_lax_intro}) which satisfies the real asymptotics~\footnote{To avoid any confusion we have used a notation for this psi-function which is different from the one used in Ref. \cite{liechty,BI03,claeys_thesis}. Indeed $\Phi_1$ and $\Phi_2$ that we use here actually correspond to $\Phi^1$ and $\Phi^2$ used in Ref. \cite{liechty,BI03,claeys_thesis}.}
\begin{eqnarray}
\Phi_1(\zeta,s) = \cos\left(\frac{4}{3} \zeta^3 + s \zeta \right) + {\cal O}(\zeta^{-1}) \;, \; \Phi_2(\zeta,s) = -\sin\left(\frac{4}{3} \zeta^3 + s \zeta \right) + {\cal O}(\zeta^{-1}) \;,
\end{eqnarray}
as $\zeta \to \pm \infty$ for $s$ real~\cite{BI03}.

The marginal pdf $P(w)$ is given by integrating the jpdf ${P}(w,s)$ over $s$.  Although the obtained explicit expression for $P(w)$ remains complicated, one can study its asymptotic behavior for $w \to \infty$ from Eqs. (\ref{explicit_P},\ref{explicit_h1_intro}) and show that
\begin{eqnarray}\label{asympt_margin_s}
P(w) = \int_{-\infty}^\infty {P}(s,w) \, ds \;, \; \log P(w) = -\frac{1}{12}w^3 + o(w^3) \;.
\end{eqnarray}     

As mentioned above, the results which we obtain here for the extreme statistics of $N$ non-intersecting Brownian excursions in the large $N$ limit allow to compute the jpdf $\hat{P}(m,t)$ of the maximum of the Airy$_2$ process minus a parabola $Y(u)$ (\ref{def_y}) and its position~\cite{spohn_praehofer,tracy_excursions,rambeau_epl,rambeau_pre}. As explained above, this in turn yields important information concerning the directed polymer with one free end (\ref{long},\ref{trans}). One expects indeed~\cite{tracy_excursions,corwin_hammond}
\begin{eqnarray}\label{jpdf_y}
\hat{P}(m,t) = \alpha \beta \, {P}(\alpha m \,, \, \beta t) \;,
\end{eqnarray}
and we show that $\alpha = 2^{2/3}$, $\beta = 2^{4/3}$. From Eq. (\ref{asympt_margin_s}), together with (\ref{jpdf_y}), one finds that the marginal pdf of the position (\ref{trans}) $\hat P(t)$ decays like 
\begin{eqnarray}\label{margin_airy}
\hat P(t) \sim \exp{\left(- \frac{4}{3}t^3\right)} \;, \; t \gg 1\;. 
\end{eqnarray}
This result (\ref{margin_airy}) establishes on firmer grounds a long standing conjecture~\cite{halpin_review}, based on a scaling argument\footnote{We thank J. Krug and J. Rambeau for pointing out these references~\cite{halpin_pra,goldschmidt}.} and on the study of an approximate model, the so-called "toy-model"~\cite{BO90} [where the Airy$_2$ process in $Y(u)$ (\ref{def_y}) is replaced by a Brownian motion (see also Refs. \cite{pld_monthus,groeneboom})], yielding $\log \hat P(t) \propto -t^3$. This behavior is also consistent with numerical studies~\cite{rambeau_epl,halpin_pra,goldschmidt}. 

As it is well known, directed polymer models as in Fig. \ref{fig_polymer} can be mapped onto stochastic growth processes in the KPZ universality classes in $1+1$ dimensions \cite{halpin_review,krug_review}. It was shown in particular~\cite{spohn_praehofer_prl,spohn_praehofer} that the fluctuations of the height field, in such processes, are governed by ${\cal F}_2$ in curved (or droplet-like) geometry, and by ${\cal F}_1$ in flat geometry. Recently, these theoretical predictions for KPZ interfaces have been observed in remarkable experiments on turbulent liquid crystals, both in curved and in flat geometries \cite{kazz_prl,kazz_nature}. The quantities which we focus on here (\ref{long}, \ref{trans}) have also a clear physical meaning for growth processes in the droplet geometry~\cite{rambeau_epl,rambeau_pre}: the maximum of $Y(u)$ corresponds to the maximal height of the droplet, while its position corresponds simply to the position of the maximal height of the droplet. Very recently, the distribution of the maximal height of this droplet was measured in the same experiment on turbulent liquid crystals and a very nice agreement with the Tracy-Widom distribution ${\cal F}_1$ [see Eq. (\ref{asympt_FN})] was found~\cite{kazz_evs}, as expected from Eqs. (\ref{scaling_eopt}, \ref{long}). 

The paper is organized as follows. In section 2, we give an expression of the jpdf of $M$ and $\tau_M$ for $N$ excursions (see Fig. \ref{fig_excursions}) in terms of discrete orthogonal polynomials. In section 3, we analyze the large $N$ and large $M$ limit, when $M$ is much larger than its mean value, $M \gg \sqrt{2N}$. In section 4 we analyze the system of orthogonal polynomials in the double scaling limit, which allows to compute the limiting jpdf $P(s,w)$ (\ref{result_intro}, \ref{explicit_P}), before we conclude in section 5. We have left in Appendix A the detailed analysis of the tail of the marginal pdf $P(w)$.

\section{The joint probability density function in terms of orthogonal polynomials}

In this section we derive an expression of the jpdf $P_N(M, \tau_M)$ in terms of discrete orthogonal polynomials, which turns out
to be very useful to perform the large $N$ analysis. The starting point of our analysis is an exact expression, using path integral for free fermions \cite{schehr_prl,degennes}, for the jpdf $P_N(M, \tau_M)$ of $M$ and $\tau_M$ given by \cite{rambeau_epl,rambeau_pre}
\begin{eqnarray}
\label{start_expr_first}
P_{N}(M,\tau_M)=\frac{A_{N}}{2^{N+1}M^{N(2N+1)+3}} && \sum_{\gras{n},n'_N} \Bigg[
(-1)^{n_N+n'_N} \ n_N^2 {n'_N}^{2}  \prod_{i=1}^{N-1} n_i^2 
\, \Delta_N(n_1^2,\dots,n_{N-1}^2, n_N^2)  \\
&& \times \Delta_N(n_1^2,\dots,n_{N-1}^2, {n'_N}^2)  \, e^{-\frac{\pi^2}{2 M^2} \sum\limits_{i=1}^{N-1} n_i^2}  e^{-\frac{\pi^2}{2M^2} \left[ (1-\tau_M) {n'_N}^2 + \tau_M n_N^2 \right]} \Bigg]\;, \nonumber
\end{eqnarray}
where $\Delta_N(\lambda_1,\dots,\lambda_N)$ is the $N\times N$ Vandermonde determinant, and where we use the notations $\gras{n}=(n_1,\dots,n_N)$ and $\sum_n \equiv \sum_{n=-\infty}^\infty$. The numerical constant $A_{N}$ is given by \cite{rambeau_pre}: 
\begin{equation}
\label{Nexcursions_normalization}
A_{N}=\frac{N \pi^{2N^2+N+2}}{2^{N^2-N/2} \prod_{j=0}^{N-1} \Gamma(2+j) \Gamma\left(\frac{3}{2}+j \right)} \;,
\end{equation}
which ensures the normalization of $P_N(M, \tau_M)$~\cite{rambeau_pre}, {\it i.e.}
\begin{eqnarray}\label{normalization}
\int_{0}^\infty dM \int_{0}^1 d\tau_M P_N(M, \tau_M) = 1 \;.
\end{eqnarray}

To study this multiple sum (\ref{start_expr_first}) we introduce discrete orthogonal polynomials $p_k(n)$ such that~\cite{GM94}
\begin{eqnarray}\label{ortho_condition}
\sum_{n=-\infty}^\infty p_k(n) p_{k'}(n) e^{-\frac{\pi^2}{2M^2} n^2} = \delta_{k,k'} h_k \;, 
\end{eqnarray}
where $p_k(n)$'s are monic polynomials of degree $k$:
\begin{eqnarray}
p_k(n) = n^k + \cdots \;,
\end{eqnarray}
and $h_k$'s are positive constants. In particular one has
\begin{eqnarray}\label{expand_vdm}
\prod_{i=1}^N n_i \, \Delta(n_1^2, n_2^2, \cdots, n_N^2) = \det_{1\leq i,j \leq N} p_{2i-1} (n_j) = \sum_{\sigma \in {\cal S}_N} \epsilon(\sigma) \prod_{i=1}^N p_{2 \sigma(i)-1}(n_i)\;,
\end{eqnarray}
where ${\cal S}_N$ is the group of permutations of size $N$ and $\epsilon(\sigma)$ is the signature of the permutation $\sigma \in {\cal S}_N$. After some manipulations, using this expansion (\ref{expand_vdm}) together with the orthogonality condition (\ref{ortho_condition}) we arrive at the following formula for $P_N(M, \tau_M)$:
\begin{eqnarray}\label{formula_inter}
P_N(M, \tau_M) = &&\frac{(N-1)! A_{N}}{2^{N+1} M^{2N^2+N+3}} \prod_{j=1}^{N} h_{2j-1} \nonumber \\
&\times& \sum_{n, m} (-1)^{n + m} n\, m \sum_{i=1}^N \frac{p_{2i-1}(n) p_{2i-1}(m)}{h_{2i-1}} e^{-\frac{\pi^2}{2M^2} \left[ (1-\tau_M) {n}^2 + \tau_M m^2 \right]} \;.
\end{eqnarray}
If one defines $\tau_M = \frac{1}{2} + u$ one finds $P_N(M,\tau_M) \equiv P_N(M,u)$ with
\begin{eqnarray}
&&P_N(M, u) = \frac{(N-1)! A_{N}}{2^{N+1} M^{2N^2+N+3}} \prod_{j=1}^{N} h_{2j-1} \sum_{k=1}^N G_{2k-1}( M, u) G_{2k-1}(M,-u) \label{start_expr} \\
&&
G_{2k-1}(M, u) = \sum_{n=-\infty}^\infty (-1)^n \, n \, \psi_{2k-1}(n) e^{- \frac{u \pi^2}{2 M^2} n^2 } \;, \label{def_G_psi}
\end{eqnarray}
where we have used the standard notation $\psi_k(n)$ for the "wave function":
\begin{eqnarray}\label{def_psi}
\psi_{k}(n) = \frac{p_k(n)}{\sqrt{h_k}} e^{-\frac{\pi^2}{4M^2} n^2} \;.
\end{eqnarray}
Using the result of Ref. \cite{forrester_npb}, one notices that $P_N(M, \tau_M = 1/2 +u)$ in Eq. (\ref{start_expr}) can be rewritten as
\begin{eqnarray}
&&P_N(M, u) =  F_N(M) \, \frac{\pi^2}{2 M^3} \sum_{k=1}^N G_{2k-1}(M, u) G_{2k-1}(M,-u) \;, \label{starting_explicit} \\
&& F_N(M) = \Pr\left[\max_{0\leq \tau \leq 1} x_N(\tau) \leq M \right] = \frac{N!}{\prod_{j=0}^{N-1} \Gamma(2+j)\Gamma(\frac{3}{2}+j)} \frac{\pi^{2N^2+N}}{2^{N^2+\frac{N}{2}} M^{2N^2+N}} \prod_{i=1}^{N} h_{2i-1} \;, 
\end{eqnarray}
where $F_N(M)$ is thus the cumulative distribution of the maximal height of $N$ non-intersecting excursions, on the unit time interval. In Ref. \cite{forrester_npb} the large $N$ asymptotic of $F_N(M)$ was carried out and it was shown that (see Ref. \cite{liechty} for a rigorous proof of this result)
\begin{equation}\label{asympt_FN}
 \lim_{N \to \infty} {F}_N \Big ( \sqrt{2N}(1 + s/(2^{7/3} N^{2/3})  \Big )= \mathcal F_1(s) \;,
 \end{equation}
 where ${\mathcal F}_1(s)$ is the Tracy-Widom distribution for $\beta = 1$ (\ref{F1_intro}). We recall that ${\cal F}_1$ admits the following explicit expression~\cite{TW96}
 \begin{eqnarray}\label{F1_intro}
 \mathcal{F}_1(s) = \exp\bigg( -\frac{1}{2} \int_s^\infty \left( 
\left(t-s\right)q^2(t)+q(t)\right)\,dt  \bigg) \;,
 \end{eqnarray}
 in terms of $q(t)$, which is the Hastings-McLeod solution to the Painlev\'e II equation (\ref{qdiff}). Note that 
 ${\cal F}_1(s)$ can also be expressed as a Fredholm determinant as~\cite{ferrari_spohn}
\begin{eqnarray}\label{f1_fredholm}
{\cal F}_1(s) = \det (I - \Pi_0 {\cal B}_s \Pi_0) \;,
\end{eqnarray}
where ${\cal B}_s$ is an integral operator with kernel
\begin{eqnarray}\label{kernel_ai1}
{\cal B}_s(x,y) = {\rm Ai}(x+y+s) \;,
\end{eqnarray}
where ${\rm Ai}$ is the Airy function and $\Pi_0$ is the projector onto the interval $[0, +\infty)$. 

The main result of this section is that we have reduced the analysis of the large $N$ analysis of $P_N(M, \tau_M)$ to the analysis of the sum over $k$ in Eq. (\ref{starting_explicit}) which involves the quantity $G_{2k-1}(M,u)$ in Eq. (\ref{def_G_psi}).

 \section{Analysis of $G_{2k-1}(M,u)$ for large $M$: large deviation analysis}

In this section we study in detail the behavior of $G_{2k-1}(M,u)$ in the limit of large $M$. More precisely, we assume here that $M \gg \sqrt{2N}$, and more precisely
\begin{eqnarray}\label{def_right_tail}
M - \sqrt{2N} = {\cal O}(\sqrt{N}) \;,
\end{eqnarray}
which corresponds to the "right tail" of the distribution of the maximum~(\ref{asympt_FN}). In this limit where $M$ is large, the discrete sum which defines the orthogonal polynomials $p_k$ in Eq. (\ref{ortho_condition}) can be replaced by an integral and therefore the orthogonal polynomials $p_k$  
are well approximated by Hermite polynomials~\cite{GM94}. For $M \gg \sqrt{2N}$ one thus has
\begin{eqnarray}\label{asympt_p}
p_k(x) = \left(\frac{M}{\sqrt{2} \pi}\right)^k H_k\left(\frac{\pi}{\sqrt{2} M}x \right) + {\cal O}(e^{-2M^2})\;,
\end{eqnarray} 
where $H_k$ is the Hermite polynomial of order $k$~\cite{szego}
\begin{eqnarray}
H_k(z) = k ! \sum_{m=0}^{\lfloor \frac{k}{2} \rfloor} \frac{(-1)^m (2z)^{k-2m}}{m! (k-2m)!} \;,
\end{eqnarray}
where $\lfloor x \rfloor$ is the largest integer not greater than $x$, while the amplitude $h_k$ in (\ref{ortho_condition}) is given by (see Ref.~\cite{GM94})
\begin{eqnarray}\label{asympt_h}
h_k =  \sqrt{2 \pi} k ! \left(\frac{M}{\pi} \right)^{2k+1}+ {\cal O}(e^{-2 M^2}) \;.
\end{eqnarray}
From Eqs. (\ref{asympt_p}) and (\ref{asympt_h}) one obtains that $\psi_k$ in Eq. (\ref{def_psi}) is given by
 \begin{eqnarray}\label{asympt_psi}
 \psi_k(x) = \frac{\pi^{\frac{1}{4}}}{2^{\frac{2k+1}{4}}} \frac{1}{\sqrt{ k! \, M}} H_k\left(\frac{\pi}{\sqrt{2} M}x \right) e^{-\frac{\pi^2}{4M^2} x^2} + {\cal O}(e^{-2M^2}) \;.
 \end{eqnarray}
 With this expression (\ref{asympt_psi}), it is then possible to study $G_{2k-1}(M,u)$ in the limit $M \gg \sqrt{2N}$ and $N$ large. We first consider the case $u=0$ (for illustration) and then present the analysis for arbitrary $u$. This then allows us to study the jpdf (\ref{starting_explicit}) in the large deviation regime. 
 
 \subsection{The case $u=0$}
 
To analyse $G_{2k-1}(M,u=0)$ in Eq. (\ref{def_G_psi}) it is convenient to use the Poisson summation formula to obtain
\begin{eqnarray}\label{poisson}
&&G_{2k-1}(M,u=0) = \sum_{n=-\infty}^\infty (-1)^n \, n \, \psi_{2k-1}(n) =\sum_{n=-\infty}^\infty \hat F(n)  \;, \\
&&\hat F(y) = \frac{1}{2 \pi} \int_{-\infty}^\infty dx \, e^{i \frac{x}{2}-i y x} \,  x \, \psi_{2k-1}\left(\frac{x}{2 \pi}\right) \;.
\end{eqnarray}
 Using the asymptotic behavior of $\psi_{2k-1}(x)$ in Eq. (\ref{asympt_psi}) one thus obtains that, for large $M$, $G_{2k-1}(M,u=0) \sim [\hat F(0) + \hat F(1)]/(2 \pi)$, yielding
 \begin{eqnarray}\label{G_integral}
 G_{2k-1}(M,u=0) \sim \frac{16}{\pi^{7/4}} \frac{2^{\frac{1}{4} - k}}{\sqrt{(2k-1)!}} M^{\frac{3}{2}}  \int_0^\infty dz \cos{(2 M z)} e^{-z^2} \, z \; H_{2k-1}(\sqrt{2} z) \;.
 \end{eqnarray}
 This integral can be performed by expanding the Hermite polynomial as \cite{szego}
 \begin{eqnarray}\label{expansion_hermite}
 H_{2k-1}(\sqrt{2} z) = (2k-1)! \sum_{p=0}^{k-1} (-1)^p \frac{2^{\frac{3}{2}(2k-2p-1)} }{p! (2k-2p-1)!} z^{2k-2p-1} \;,
 \end{eqnarray}
 and then integrate over $z$ term by term in Eq. (\ref{G_integral}). After some algebra one finds that $G_{2k-1}(M,u=0)$ can be rewritten in terms of Hermite polynomials as
 \begin{equation}\label{asympt_G_u0}
 G_{2k-1}(M,u=0) \sim \frac{2^{\frac{11}{4}}}{\pi^{\frac{5}{4}}} \frac{(-1)^k}{2^k \sqrt{(2k-1)!}} M^{\frac{3}{2}} e^{-M^2} \left(H_{2k}(\sqrt{2} \, M) - M \sqrt{2} H_{2k-1}(M \sqrt{2})\right)  \;.
 \end{equation}
 As we show below, this analysis can be extended to any finite value of $u$. 
 
 \subsection{The case of finite $u$}
 
 For finite $u$ the Poisson summation formula (\ref{poisson}) yields the following asymptotic expression
 \begin{equation}\label{G_integral_ufinite}
 G_{2k-1}(M,u) \sim (-1)^k \frac{16}{\pi^{7/4}} \frac{2^{\frac{1}{4} - k}}{\sqrt{(2k-1)!}}  \frac{M^{\frac{3}{2}}}{1+2u} \int_0^\infty dz \cos{\left(\frac{2M}{\sqrt{1+2u}}  z\right)} e^{-z^2} \, z \; H_{2k-1}\left(\sqrt{\frac{2}{1+2u}} z\right) \;.
 \end{equation}
 Expanding again the Hermite polynomial as before (\ref{expansion_hermite}) and integrating term by term, one obtains
 \begin{eqnarray}\label{asympt_G_ufinite}
 G_{2k-1}(M,u) &\sim&(-1)^k \frac{2^{11/4}}{\pi^{5/4}} \frac{1}{2^k \sqrt{(2k-1)!}} \frac{M^{\frac{3}{2}} e^{-\frac{M^2}{1 + 2u}}}{(1-2u)^{3/2}} \left(\frac{1-2u}{1+2u} \right)^k \\
&\times& \left[\sqrt{\frac{1-2u}{1+2u}} H_{2k}\left(M \, \sqrt{\frac{2}{1-4u^2}} \right) - \sqrt{2} M H_{2k-1}\left(M \sqrt{\frac{2}{1-4u^2}}\right)\right]  \;. \nonumber
 \end{eqnarray}
Given this expression (\ref{asympt_G_ufinite}), it appears that the sum over $k$ which enters the expression of $P_N(M, \tau_M)$ in Eq. (\ref{starting_explicit}) is actually dominated, for large $M$, by the large values of $k$, which we now study.  
 
 \subsection{Asymptotic limit of $G_{2k-1}(M,u)$ for large $M$ and large $k$}
 
 In the limit of large $M$ we are thus interested in the behavior of $G_{2k-1}(M,u)$ for large $k \sim {\cal O}(N)$ with 
 \begin{eqnarray}\label{def_c}
 c = \frac{2 k}{M^2} 
 \end{eqnarray}
 fixed. Here $c<1$, corresponding to $M \gg \sqrt{2 N}$ [see Eq. (\ref{def_right_tail})]. To study the behavior of $G_{2k-1}(M,u)$ in Eq. (\ref{asympt_G_ufinite}) in the limit of large $M$ and large $k$ it is convenient to use an integral representation of Hermite polynomials (see also Ref. \cite{GM94,nadal} for a similar calculation with different orthogonal polynomials). A convenient one is the one coming from the relation for the exponential generating function~\cite{abramowitz,szego}
\begin{eqnarray}
\sum_{p=0}^\infty H_p(x) \frac{t^p}{p!} = \exp{(2 x \,t - t^2)} \;,
\end{eqnarray}
 which allows to write
 \begin{eqnarray}\label{contour_rep}
 H_p(x) = p ! \oint \frac{dt}{2 i \pi} \frac{1}{t^{p+1}} e^{2 xt -t ^2} \;,
 \end{eqnarray}
 where the contour must encircle the value $t=0$. From this representation (\ref{contour_rep}) one obtains
 \begin{eqnarray}
 H_{2k}\left(M \, \sqrt{\frac{2}{1-4u^2}} \right) = (2k)! \oint \frac{dt}{2i \pi} \frac{1}{t}\exp{\left(2 \sqrt{\frac{2}{1-4u^2}} M t - t^2 - c \log{t} \right)} \;,
 \end{eqnarray}
 which is quite convenient for a large $M$ analysis. Performing the change of variable $t = y M$, one gets
\begin{eqnarray}
 &&H_{2k}\left(M \, \sqrt{\frac{2}{1-4u^2}} \right) = \frac{(2k)!}{M^{2k}} \oint \frac{dy}{2i \pi} \frac{1}{y}\exp{(-M^2 \phi(y))} \;, \label{def_contour} \\
 &&\phi(y) = y^2 - 2 \sqrt{\frac{2}{1-4u^2}} y + c \log{y} \;.
 \end{eqnarray}
 For $c<1$, the contour integral in (\ref{def_contour}) can be evaluated for large $M$ by a saddle point, which yields 
 \begin{eqnarray}\label{G_saddle}
 G_{2k-1}(M,u) &\sim& (-1)^k \frac{2^{11/4}}{\pi^{5/4}} \frac{1}{2^k \sqrt{(2k-1)!}} \frac{M^{\frac{3}{2}} e^{-\frac{M^2}{1 + 2u}}}{(1-2u)^{3/2}} \left(\frac{1-2u}{1+2u} \right)^k \nonumber \\
 &\times& \frac{(2k)!}{M^{2k}} e^{-M^2 \phi(y^*)} \frac{1}{\sqrt{2 \pi} \sqrt{M^2 |\phi''(y^*)|}} \left( \frac{1}{y^*}\sqrt{\frac{1-2u}{1+2u}} - \frac{\sqrt{2}}{c} \right) \;,
 \end{eqnarray}
 with
 \begin{eqnarray}
 y^* &=&  \frac{1-\sqrt{1-c\, \rho}}{\sqrt{2 \, \rho}} \;, \; \phi(y^*) = -\frac{2-2\sqrt{1-c\rho} + c\rho(1 +  \ln{(2\rho)} - 2 \ln{(1-\sqrt{1-c\rho})}) }{2 \rho} \;, \\
 \phi''(y^*)  &=& 2 - 2\frac{c\rho}{(1- \sqrt{1-c\rho})^2} < 0 \;.
 \end{eqnarray}
 in terms of 
 \begin{eqnarray}
 \rho = 1 - 4u^2 \;.
 \end{eqnarray}
 On the other hand, using Stirling's formula, one obtains, in the large $M$ limit, keeping $c=2k/M^2$ fixed~(\ref{def_c})
  \begin{eqnarray}\label{stirling}
 \frac{\sqrt{(2k)!}}{2^k M^{2k}} \sim (2 \pi c)^{1/4} \sqrt{M} e^{-M^2 (\frac{c}{2} + \frac{c}{2} \ln 2 - \frac{c}{2} \ln c)} \;.
  \end{eqnarray}
 Finally, combining Eq. (\ref{G_saddle}) together with Eq. (\ref{stirling}) one obtains
  \begin{eqnarray}\label{large_dev_G}
  G_{2k-1}(M,u) &\sim& (-1)^k\frac{2^{5/2}}{\pi^{3/2}} \frac{M^2}{(1-2u)^{3/2}} \frac{c^{3/4}}{\sqrt{|\phi''(y^*)|}}  \left( \frac{1}{y^*}\sqrt{\frac{1-2u}{1+2u}} - \frac{\sqrt{2}}{c} \right) \nonumber \\
  &\times& \exp{\left[-M^2 \left( \frac{c}{2} \left(1 - \ln {(c/2)} + \ln{\left(\frac{1+2u}{1-2u}\right)} \right) + \frac{1}{1+2u}+\phi(y^*)\right)\right]} \;.
  \end{eqnarray} 
 Let us analyse this formula for $u=0$, where $G_{2k-1}(M,u=0)$ takes a simpler form given by
 \begin{eqnarray}
 &&G_{2k-1}(M,u=0) \sim (-1)^k\frac{4}{\pi^{3/2}} M^2 \sqrt{1-\sqrt{1-c}} \left(\frac{1-c}{c}\right)^{1/4} \nonumber \\
  && \times \exp{\left[-M^2 \left(\sqrt{1-c}+c \ln \left(1-\sqrt{1-c}\right)-\frac{1}{2} c \ln c \right) \right]} \;.
  \end{eqnarray}
  In the limit where $c \to 1$, one finds
  \begin{eqnarray}\label{asympt_cto1_1}
  G_{2k-1}(M,u=0)  \sim (-1)^k \frac{4}{\pi^{3/2}} M^2 (1-c)^{1/4} e^{- \frac{2}{3} M^2 (1-c)^{3/2}} \;.
  \end{eqnarray} 
 In particular, in the regime where 
 \begin{equation}\label{c_close1}
 (1-c) = x M^{-4/3} \;,
 \end{equation}
 which corresponds precisely to the tail of the double scaling regime to be studied in section 4, one finds
 \begin{eqnarray}\label{asympt_cto1_2}
 G_{2k-1}(M,u=0) \sim (-1)^k \frac{4}{\pi^{3/2}}  M^{5/3} x^{1/4} e^{-\frac{2}{3} x^{3/2}} \sim (-1)^{k+1} \frac{8}{\pi} M^{5/3} {\rm Ai}'(x)  \;,
 \end{eqnarray}
 where ${\rm Ai}'(x)$ is the derivative of the Airy function~\cite{abramowitz} and where the last relation in Eq. (\ref{asympt_cto1_2}) holds only for large $x$. Note that this result (\ref{asympt_cto1_2}), involving ${\rm Ai}'(x)$ can also be directly obtained from the above expression in terms of Hermite polynomials given in Eq. (\ref{asympt_G_u0}). To do so, we make use of the Plancherel-Rotach formula \cite{szego,plancherel}    
 \begin{eqnarray}\label{plancherel_rotach}
 \exp{(-z^2/2)} H_{2k+m}(z) = (4k)^{\frac{m}{2}} \pi^{\frac{1}{4}} 2^{k+\frac{1}{4}} \sqrt{(2k)!} (2k)^{-\frac{1}{12}}\left( {\rm Ai}(t) - \frac{m}{(2k)^{\frac{1}{3}}} {\rm Ai}'(t) + {\cal O}(k^{-\frac{2}{3}}) \right) ,
 \end{eqnarray}
 valid for large $k$, where we have set
\begin{eqnarray}\label{def_z}
z = (4k)^{1/2} + 2^{-2/3}k^{-1/6} t \;.
\end{eqnarray} 
Indeed, if one applies this Plancherel-Rotach formula (\ref{plancherel_rotach}) to Eq. (\ref{asympt_G_u0})   with $z \to \sqrt{2} M = (4k)^{1/2} + 2^{-2/3} k^{-1/6} x$ [see Eqs. (\ref{def_c}, \ref{c_close1})], one obtains precisely the above asymptotic behavior (\ref{asympt_cto1_2}) in terms of~${\rm Ai}'(x)$. 
 
It is instructive, and useful for the forthcoming double scaling analysis, to apply this Plancherel-Rotach formula (\ref{plancherel_rotach}) to $G_{2k-1}(M,u)$ for finite $u$, as given in Eq. (\ref{asympt_G_ufinite}), where we also rescale $u$ according~to 
\begin{eqnarray}\label{def_v}
u = v \; M^{-2/3} \;.
\end{eqnarray}
One can then apply this formula (\ref{plancherel_rotach}) to (\ref{asympt_G_ufinite}) with $z  \to M \sqrt{2/(1-u^2)} = (4k)^{1/2} + (x + 4v^2) 2^{-2/3} k^{-1/6}$, which yields
\begin{eqnarray}\label{p_r_finite_u}
G_{2k-1}(M,u) \sim (-1)^{k+1}\frac{8}{\pi}e^{\frac{16 \, v^3}{3}+2 v\, x} \left(2 v {\rm Ai}(4 v^2 + x) +  {\rm Ai}'(4v^2 + x)\right)  \;.
\end{eqnarray}
Note that the same function (\ref{p_r_finite_u}), albeit with different arguments, enters the expression of the jpdf given in Ref. \cite{quastel_jpdf} [see Eqs. (\ref{expr_quastel}), (\ref{expr_quastel_phi}) below]. Here this function arises naturally as the Plancherel-Rotach asymptotic of our large $N$, large deviation, regime. This fact will allow us to identify the coefficients $\alpha$ and $\beta$ in Eq. (\ref{jpdf_y}).

 \subsection{Large deviation regime of the joint probability density function for $M \gg \sqrt{2 N}$}  
 
 From the asymptotic behavior obtained in the previous section (\ref{large_dev_G}) it is now possible to obtain the behavior of the jpdf $P_N(M, \tau_M) \equiv P_N(M, u)$ with $\tau_M = 1/2 + u$. From Eq. (\ref{starting_explicit}), using that $F_N(M) \sim 1$ when $M \gg \sqrt{2N}$ (see Ref. \cite{forrester_npb} for a more refined analysis of this right tail) one has
\begin{eqnarray}\label{large_dev_first_step}
P_N(M, \tau_M) & \sim& \frac{\pi^2}{2 M^3} \sum_{k=1}^N G_{2k-1}(M, u) G_{2k-1}(M,-u) \\
&\sim& \frac{2^4 M}{\pi} \sum_{k=1}^N \frac{M^4}{(1-4u^2)^{3/2}} \frac{1}{|\phi''(y^*)|}  \frac{2(1-c)}{c^{1/2}} \exp{\left[-M^2 \varphi(c,u)\right]} \;,\; c = \frac{2k}{M^2} \\
\varphi(c,u) &=& 2  \left( \frac{c}{2} \left(1 - \ln {(c/2)}  \right) + \frac{1}{1-4u^2}+\phi(y^*)\right) \\
&=& \frac{2 \sqrt{1-c\rho}}{\rho} - c \ln{(c\rho)} + 2 c \ln{(1 - \sqrt{1-c\rho})} \;, \; \rho = 1 - 4u^2 \;. \label{def_psi_large_dev}
 \end{eqnarray}
 One can easily check that for any value of $u \in [-1/2, 1/2]$, $\varphi(c,u)$ is a decreasing function of $c$. Therefore the sum over $k$ in Eq. (\ref{large_dev_first_step}) is dominated, for large $M$, by $k = N$. This yields, in this regime (in the sense of logarithmic equivalent, with $2N/M^2$ fixed)
 \begin{eqnarray}
 P_N(M,u) \sim \exp{\left[- M^2 \varphi(2N/M^2, u) \right]} \;.
 \end{eqnarray} 
In particular, when $M^2 \to 2 N$, one has 
\begin{eqnarray}
&&P_N(M,u) \sim \exp{\left(- 2 N \varphi(1, u)\right)} \;, \\
&&\varphi(1,u) = \frac{4 |u|}{1-4u^2} - \ln{(1-4u^2)} + 2 \ln{(1-2 |u|)} = \frac{32}{3}|u|^3 + {\cal O}(|u|^5)  \;.
\end{eqnarray}   
This cubic behavior for small $u$ implies 
\begin{eqnarray}\label{cubic_largedev}
P_N(M \sim \sqrt{2N},u  \sim v M^{-1/3}) \sim \exp{\left(-\frac{32}{3}|v|^3\right)} \;,
\end{eqnarray}
 which exhibits an interesting non-Gaussian behavior. We will recover this cubic behavior (\ref{cubic_largedev}) below, in the double scaling.

 \section{Double scaling regime}
 
 The above results concern large deviations: they describe the fluctuations for which $M$ is much bigger than the mean value, when $M \gg \sqrt{2N}$. However, it was shown in Ref. \cite{forrester_npb} that the typical fluctuations of $M$ behave like $M - \sqrt{2N} = {\cal O}(N^{-1/6})$. In the language of Random Matrix Theory, this corresponds to a double scaling regime. We thus set $M - \sqrt{2N} = s N^{-1/6}$ where $s$ is fixed and we are led to study the quantity $G_{2k-1}(M,u)$ where both $k \sim N$ and $M^2 \sim 2N$. To study this regime, we need to study the discrete orthogonal polynomials $p_k$'s (\ref{ortho_condition}) beyond the approximation where $p_k$'s are estimated by Hermite polynomials (\ref{asympt_p}). To perform this analysis, we study in detail the three terms recursion relation satisfied by these orthogonal polynomials \cite{GM94}. This recursion relation for $p_k$'s yields differential recursion relations for $G_{2k-1}(M,u)$, which we derive in the first subsection. In the second subsection,  we remind some useful results obtained by Gross and Matytsin \cite{GM94} concerning the analysis of this three terms recurrence in the double scaling limit. In the third subsection we obtain a differential equation and a partial differential equation satisfied by $G_{2k-1}(M,u)$ in the double scaling limit and obtain the limiting form ${P}(s,w)$
 of the jpdf $P_N(M,\tau_M)$ for large $N$. In the fourth subsection, we show how to solve the aforementioned differential equations. In the last subsection, we present an asymptotic analysis of the marginal pdf ${P}(w)$.   
 
\subsection{Differential Recursion Relations}

In this subsection, we derive two distinct differential recursion relations satisfied by $G_{2k-1}(M,u)$ which will be very useful to perform the large $N$ analysis of the expression above (\ref{start_expr}). For this purpose we remind that the discrete orthogonal polynomials $p_k$'s defined by the orthogonality relation (\ref{ortho_condition}) satisfy the recursion relation \cite{szego,mehta_book,forrester_book}
\begin{eqnarray}\label{recurrence_p}
x \, p_{k} (x) = p_{k+1}(x) + R_k \; p_{k-1}(x) \;, \; R_k = \frac{h_k}{h_{k-1}} \;,
\end{eqnarray}
such that $\psi_k$'s in Eq. (\ref{def_psi}) satisfy
\begin{eqnarray}\label{recurrence_psi}
x \psi_k(x) =  \gamma_{k+1} \psi_{k+1}(x) + \gamma_k \psi_{k-1}(x) \;, \; \gamma_k = \sqrt{R_k} = \sqrt{\frac{h_k}{h_{k-1}}} \;.
\end{eqnarray}
From this recursion relation (\ref{recurrence_psi}) one deduces easily the following differential recursion relation
\begin{equation}
\frac{\partial}{\partial u} G_{2k-1}= -\frac{\pi^2}{2 M^2} \left[\gamma_{2k} \gamma_{2k+1} G_{2k+1} + (\gamma^2_{2k} + \gamma^2_{2k-1}) G_{2k-1} + \gamma_{2k-2} \gamma_{2k-1} G_{2k-3} \right] \label{ed_u} \;,
\end{equation}
 where, for clarity, we used the notation $G_{2k-1} \equiv G_{2k-1}(M,u)$. 
 
One can also derive a differential recursion relation for $G_{2k-1}(M,u)$ when $M$ is varied. By differentiating the orthogonality relation (\ref{ortho_condition}) with respect to $M$, one obtains the identity (see also Ref.~\cite{eynard_harnad} for a similar relation for more general matrix models):
\begin{eqnarray}
-\frac{M^3}{2\pi^2} \frac{\partial}{\partial M} \psi_k = \frac{1}{4} \gamma_k \gamma_{k-1} \psi_{k-2} - \frac{1}{4} \gamma_{k+2} \gamma_{k+1} \psi_{k+2} \;,
\end{eqnarray}
from which one obtains straightforwardly:
\begin{eqnarray}
-\frac{M^3}{2 \pi^2}\frac{\partial}{\partial M} G_{2k-1}&=&  \Bigg[\left(\frac{1}{4}-\frac{u}{2}\right)\gamma_{2k-1} \gamma_{2k-2} G_{2k-3} - \frac{u}{2} \left(\gamma^2_{2k} + \gamma^2_{2k-1} \right)G_{2k-1} \nonumber \\
&-& \left(\frac{1}{4} + \frac{u}{2} \right) \gamma_{2k} \gamma_{2k+1} G_{2k+1} \Bigg] \;. \label{ed_M}
\end{eqnarray} 
These two differential recursion relations (\ref{ed_u}, \ref{ed_M}) will be analyzed below in the double scaling limit where they will lead, respectively, to a partial differential equation and an ordinary differential equation for $G_{2k-1}$.

 \subsection{Three terms recursion relation in the double scaling limit}
 
 In this regime, it was shown by Gross and Matytsin \cite{GM94} and later, rigorously, by Liechty \cite{liechty}, that the coefficients $R_k$ in Eq. (\ref{recurrence_p}) take the following scaling form:
\begin{eqnarray}\label{R_dble_scaling}
&&R_{2k} = \frac{M^4}{\pi^2} - M^{10/3} f_1(x_{2k}) + M^{8/3} f_2^+ (x_{2k}) + {\cal O}(M^2) \;, \; x_{2k} = M^{4/3}\left(1-\frac{2k}{M^2} \right) \;, \\
&&R_{2k+1} = \frac{M^4}{\pi^2} + M^{10/3} f_1(x_{2k+1}) + M^{8/3} f_2^-(x_{2k+1}) + {\cal O}(M^2)  \;, \; x_{2k+1} = M^{4/3}\left(1-\frac{2k+1}{M^2}  \right),
\end{eqnarray}
where the function $f_1$ satisfies a Painlev\'e II equation (PII), corresponding to $\alpha = 0$,
\begin{eqnarray}\label{eq_f1}
f_1''(x) = 4 x f_1(x) + \frac{\pi^2}{2} f_1^3(x) \;, \; f_1(x) \mathop{\sim}\limits_{x\rightarrow \infty}  - \frac{2^{5/3}}{\pi^2}{\rm Ai}(2^{2/3} x) \;,
\end{eqnarray}
where ${\rm Ai}(y)$ is the standard Airy function. Note the minus sign in the asymptotic behavior wich is missing in Ref. \cite{GM94}. It can be expressed in terms of the Hastings-McLeod solution of PII given in Eq.~(\ref{qdiff}) as
\begin{eqnarray}\label{rel_f1_mcleod}
f_1(x) = -\frac{2^{5/3}}{\pi^2} q(2^{2/3} x) \;.
\end{eqnarray} 
One can also show \cite{GM94} that the functions $f_2^+, f_2^-$ in Eq. (\ref{R_dble_scaling}) satisfy
\begin{eqnarray}
f_2^+(x) + f_2^-(x) = - \frac{2}{\pi^2} x + \frac{\pi^2}{2} f_1^2(x) \;.
\end{eqnarray}
We also notice that $f_2^+ = f_2^{-}$ while this identity is not really needed here (in fact one can check that this is always the combination $f_2^+ + f_2^-$ which enters into the calculations \cite{satya_private}).

\subsection{Recursion relations for $G_{2k-1}(M,u)$ in the double scaling limit}
 
 From the large deviation analysis performed above and in particular from the behavior obtained in Eqs. (\ref{asympt_cto1_2}), (\ref{p_r_finite_u}) in terms of the variables $x$ in (\ref{c_close1}) and $v$ in (\ref{def_v}), one also expects that $G_{2k-1}(M,u)$ will be, in the double scaling limit, a function of the variables
 \begin{eqnarray}\label{double_variable}
 x_{2k} = M^{4/3}\left(1- \frac{2k}{M^2} \right) \;, \; v = u M^{2/3} \;.
 \end{eqnarray}  
On the other hand, due to the term $(-1)^{k+1}$ in (\ref{p_r_finite_u}), one expects that that $G_{2k-1}(M,u)$ behaves differently for $k$ odd or $k$ even, also in the double scaling limit~\cite{GM94}. Therefore, guided by this analysis, and by the behavior of $R_k$ in Eq. (\ref{R_dble_scaling}), one assumes the following ansatz
\begin{eqnarray}\label{ansatz_g_double_scaling}
&&G_{2k-1}(M,u) = M^{5/3} \left(g_1^+(x_{2k}, v) + M^{-2/3} g_2^+(x_{2k},v) + M^{-4/3} g_3^+(x_{2k},v) + {\cal O}(M^{-2})\right), k \; {\rm even} \\
&&G_{2k-1}(M,u) = M^{5/3} \left(g_1^-(x_{2k}, v) + M^{-2/3} g_2^-(x_{2k},v) + M^{-4/3} g_3^-(x_{2k},v) + {\cal O}(M^{-2})\right), k \; {\rm odd},
\end{eqnarray}
where $g_1^{\pm}, g_2^{\pm}, g_3^{\pm}$ are some functions which remain to be determined. From Eq. (\ref{R_dble_scaling}) one has 
 \begin{eqnarray}
 &&\gamma_{2k} \gamma_{2k+1} = \sqrt{R_{2k} R_{2k+1}}= \frac{M^4}{\pi^2} - M^{8/3}\left(\frac{\pi^2}{4}f_1^2(x_{2k}) + \frac{1}{2} f_1'(x_{2k})+\frac{1}{\pi^2} x_{2k} \right) + {\cal O}(M^2) \;, \label{prod_gamma} \\
 &&\gamma^2_{2k} + \gamma^2_{2k-1} = R_{2k} + R_{2k-1} = 2 \frac{M^4}{\pi^2} + M^{8/3} \left(\frac{\pi^2}{2}f_1^2(x_{2k}) + f_1'(x_{2k})-\frac{2}{\pi^2} x_{2k} \right) + {\cal O}(M^2) \;. 
 \end{eqnarray}
 
 \subsubsection{The case $u=0$}
 
 Substituting this ansatz (\ref{ansatz_g_double_scaling}) into Eq. (\ref{ed_M}), for $u=0$, and performing the same analysis as the one done in Ref. \cite{GM94}
 one obtains 
 \begin{eqnarray}
 g_1^+(x,0) = - g_1^-(x,0) = g_1(x,0) \;,
 \end{eqnarray}
 where $g_1(x,0)$ is the unique solution of the third order linear differential equation 
 \begin{eqnarray}
 &&\frac{4}{3\pi^2} y'''(x) + y'(x) \left[\frac{2x}{3\pi^2} - 8 u_1(x) \right] - y(x) \left[\frac{5}{3\pi^2}+ 4u'_1(x) \right] = 0 \;, \label{3rd_lde} \\
 && u_1(x) = \frac{1}{4}\left[\frac{1}{2} f_1'(x) + \frac{\pi^2}{4} f_1^2(x) + \frac{x}{\pi^2} \right] \;, \label{def_u1}
 \end{eqnarray} 
whose large $x$ behavior has to match the large deviation regime, where $M \to \sqrt{2N}$ from above [see Eq. (\ref{asympt_cto1_2})]
 \begin{eqnarray}
 g_1(x,0) \sim -\frac{8}{\pi} {\rm Ai}'(x) \;, \; x \to + \infty \;.
 \end{eqnarray}
 As a check, if one uses the large $x$ behavior of $u_1(x)$ in Eq. (\ref{def_u1}), $u_1(x) \sim x/(4 \pi^2)$ one finds that the above third order differential equation reads (\ref{3rd_lde}), for large $x$
 \begin{eqnarray}
 y'''(x) + x\, y'(x)  - 2 y(x) = 0 \;, \label{lde_asympt}
 \end{eqnarray}
 which admits $y(x) = {\rm Ai}'(x)$ as one of its three independent solutions. It is convenient to write this differential equation (\ref{3rd_lde}) in terms of the Hastings-McLeod solution $q(s)$, which is related to $f_1(x)$ through the above relation (\ref{rel_f1_mcleod}). In particular one finds straightforwardly
 \begin{eqnarray}
  u_1(x) = \frac{1}{2^{2/3} \pi^2} (q^2(s)-q'(s)+ \frac{s}{4}) \;, \; s = 2^{2/3} x\;.
 \end{eqnarray}
From Eq. (\ref{3rd_lde}) one has also
 \begin{eqnarray}
 g_1(x,v) = f(s = 2^{2/3} x,w = 2^{7/3} v) \;,
 \end{eqnarray}
 where $f(s,w=0)$ is the solution of 
 \begin{eqnarray}\label{third_order_mcleod}
 4 y'''(s) - y'(s)\left[6\left(q^2-q'\right)+s\right] - y(s)\left[3\left(q^2-q'\right)' + 2\right] = 0 \;,
 \end{eqnarray}
 which behaves, for large $s$, like
 \begin{eqnarray}
 f(s,w=0) \sim -\frac{8}{\pi} {\rm Ai}'(2^{-2/3} s) \;, \; s \to \infty \;.
 \end{eqnarray}

 \subsubsection{The case $u \neq 0$}
 
 For $u \neq 0$ one obtains that $g_1(x,v)$, with $u=vM^{-2/3}$ is given by the solution of the following differential equation where $v$ plays the role of 
 a simple parameter
 \begin{eqnarray}
 \frac{4}{3\pi^2} y_v'''(x) - v \frac{16}{3 \pi^2}y_v''(x) &+& y_v'(x) \left[\frac{2x}{3\pi^2} - 8 u_1(x) \right] \\
 &&- y_v(x) \left[\frac{5}{3\pi^2}+ 4u'_1(x)-\frac{4v}{3}(\pi^2 f_1^2(x) + 2f_1'(x)) \right] = 0 \;,  \label{3rd_lde_u} 
\end{eqnarray}
whose large $x$ behavior is given by Eq. (\ref{p_r_finite_u}):
\begin{eqnarray}
g_1(x,v) \sim -\frac{8}{\pi}e^{\frac{16 \, v^3}{3}+2 v\, x} \left(2 v {\rm Ai}(4 v^2 + x) +  {\rm Ai}'(4v^2 + x)\right) \;.
\end{eqnarray}
We now perform the change of variable
\begin{eqnarray}
 g_1(x,v) = f(s = 2^{2/3} x,w = 2^{7/3} v) \;,
 \end{eqnarray}
and one then obtains, from Eq. (\ref{3rd_lde_u}),  an equation satisfied by $f(s,w)$ 
\begin{eqnarray}\label{eq_y_s_w}
4y_w''' - 2 w y_w'' - y'_w\left[6(q^2-q') + s\right] - y_w\left[3(q^2-q')' +2- 2 w(q^2-q')  \right] = 0  \;,
\end{eqnarray} 
where $y_w \equiv y_w(s)$ and where $f(s,w)$ is given by the unique solution of Eq. (\ref{eq_y_s_w})
which, for large $s$, behaves as
\begin{eqnarray}
f(s,w) \sim \frac{-2^{11/3}}{\pi}e^{\frac{1}{24}w^3+\frac{w\, s}{4}} \left(\frac{w}{4}{\rm Ai}(w^2/2^{8/3} + s/2^{2/3}) + \frac{1}{2^{2/3}} {\rm Ai}'(w^2/2^{8/3} + s/2^{2/3}) \right) \;. \label{asympt_h1}
\end{eqnarray}

 On the other hand, from the recursion relation (\ref{ed_u}), one obtains, in the double scaling limit, that $f(s,w)$ satisfies the following equation
 \begin{eqnarray}\label{schrod_eq}
\frac{\partial}{\partial w} f(s,w) = \left[\frac{\partial^2}{\partial s^2} - (q^2- q') \right] f(s,w)  \;,
 \end{eqnarray}
 which has actually the structure of a supersymmetric Schr\"odinger equation. This in turns means that $f(s,w)$ satisfies a Fokker-Planck equation. The connection between the Airy$_2$ process and this diffusion process deserves certainly to be explored. This equation (\ref{schrod_eq}), together with the third order differential equation satisfied by $f(s,w=0)$ (\ref{eq_y_s_w}) determines uniquely the function $f(s,w)$.  
 \subsubsection{Limiting form of the jpdf in the large $N$ limit}
 
 We recall that the jpdf $P_N(M,\tau_M)$ is given by
 \begin{eqnarray}
 &&P_N(M, \tau_M) =  F_N(M) \, \frac{\pi^2}{2 M^3} \sum_{k=1}^N G_{2k-1}(M, u) G_{2k-1}(M,-u) \;, \; \tau_M = \frac{1}{2} + u \;, 
 \end{eqnarray}
 using the scaling form found above (\ref{ansatz_g_double_scaling}), one finds
 \begin{eqnarray}
&&P_N(M, \tau_M) = F_N(M) \frac{\pi^2}{2} M^{1/3} \\
&&\times \sum_{k=1}^N g_1\left(\frac{M^2-2k}{M^{2/3}},(\tau_M-1/2) M^{2/3}\right) g_1\left(\frac{M^2-2k}{M^{2/3}},-(\tau_M-1/2)M^{2/3}\right)  + {\cal O}(M^{-2/3})\;. \nonumber  
 \end{eqnarray}
 If we set 
 \begin{eqnarray}
 &&M = \sqrt{2N} + 2^{-11/6} N^{-1/6} s \; , \; {\it i.e.} \; \;  M^2 = 2 N + 2^{-1/3} N^{1/3} s + {\cal O}(N^{-1/3})  \;, \nonumber \\
 && \tau_M - \frac{1}{2} = 2^{-7/3} w M^{-2/3} = 2^{-8/3} w N^{-1/3} \;,
 \end{eqnarray}
 one finds that for large $N$, using also the result above (\ref{asympt_FN}), obtained in Ref. \cite{forrester_npb}
 \begin{eqnarray}
 &&P_N(\sqrt{2N} + 2^{-11/6} N^{-1/6} s, 2^{-8/3} N^{-1/3} w) \\
 &&\sim 2^{9/2} N^{1/2} \frac{\pi^2}{2^{16/3}} {\cal F}_1(s)  \frac{1}{N^{1/3}} \sum_{j=1}^{N} f\left(2^{4/3} \frac{j}{N^{1/3}} + s,w\right)  f\left(2^{4/3} \frac{j}{N^{1/3}} + s,-w\right) \;, \nonumber
\end{eqnarray}
where we have simply made the change of variable $j = N-k$. In the large $N$ limit the discrete sum over $j$ converges to an integral where, thanks to the behavior of $f(x,w)$ for large $x$ the upper bound of the integral can be safely taken to infinity. Finally one obtains the large $N$ limit as
\begin{eqnarray}
\lim_{N \to \infty} 2^{-\frac{9}{2}}N^{-\frac{1}{2}} P_N(\sqrt{2N} + 2^{-\frac{11}{6}}\, s \,N^{-\frac{1}{6}}, \frac{1}{2} + 2^{-\frac{8}{3}} \, w \, N^{-\frac{1}{3}} ) = {P}(s,w) \;, \label{convergence}
\end{eqnarray}
where $P(s,w)$ is given by
\begin{eqnarray}
P(s,w) = \frac{\pi^2}{2^{20/3}} {\cal F}_1(s) \int_{s}^\infty f(x,w) f(x,-w) \, dx \;, \label{joint_explicit}
\end{eqnarray}
where $f(x,w)$ is the unique solution of the third order differential equation in Eq. (\ref{eq_y_s_w}) which has the asymptotic behavior given in Eq. (\ref{asympt_h1}) for large $x$. It also satisfies the Schr\"odinger-like equation in Eq.~(\ref{schrod_eq}). In the following we show how to solve this equation (\ref{eq_y_s_w}) in terms of a psi-function associated to the Painlev\'e-II equation. This second equation (\ref{schrod_eq}) turns out to be very useful to guess the form of the solution.

 \subsection{Solution of equations (\ref{third_order_mcleod}) and (\ref{schrod_eq}) in terms of psi-function associated to Painlev\'e-II}

% {\bf Mention that such ode with Painlevé transcendents have appeared in the lit....}
 
 Such differential equations (\ref{eq_y_s_w}, \ref{schrod_eq}) where the coefficients involve Painlev\'e II transcendents have appeared several times in the literature on related subjects (see for instance \cite{baik_rains}, \cite{praehofer_spohn}, \cite{forrester_spike}). In these cases it was useful to introduce Lax pairs associated to the Painlev\'e II equation. On the other hand, in Ref. \cite{liechty} it was shown that the Christoffel-Darboux kernel associated to the orthogonal polynomials studied here (\ref{ortho_condition}) is described,  in the double scaling limit, in terms of psi-functions related to a Lax pair for the Painlev\'e II equation. This gives a hint that this Lax pair might be relevant to analyze these differential equations~(\ref{eq_y_s_w}, \ref{schrod_eq}). We introduce briefly this Lax pair and refer 
 the reader to Ref.~\cite{liechty,claeys_thesis} for more detail. Following Ref. \cite{BI03}, one considers the linear differential equations for a 2-vector $\Psi \equiv \Psi(\zeta,s)$,
 \begin{eqnarray}\label{system_lax}
 \frac{\partial}{\partial \zeta} \Psi = A \Psi \;, \; \frac{\partial}{\partial s} \Psi = B \Psi \;,
 \end{eqnarray} 
 where the $2 \times 2$ matrices $A \equiv A(\zeta, s)$ and $B \equiv B(\zeta,s)$ are given by
 \begin{eqnarray}\label{matrice_A}
 A(\zeta,s) = \left( 
 \begin{array}{c c}
 4 \zeta q &  4 \zeta^2 + s + 2q^2 + 2r\\
 -4 \zeta^2 - s - 2 q^2 + 2r & -4 \zeta q
 \end{array}\right) 
 \end{eqnarray}
 and
 \begin{eqnarray}\label{matrice_B}
 B(\zeta,s) = \left( 
 \begin{array}{c c}
 q &  \zeta \\
 -\zeta & - q
 \end{array}\right) \;.
 \end{eqnarray}
 These matrices $A$ and $B$ constitute a Lax pair associated to the Hastings-McLeod solution of the Painlev\'e II equation: this means
 that the compatibility equation of this system (\ref{system_lax}), {\it i.e.} $\partial_s \partial_\zeta \Psi = \partial_\zeta \partial_s \Psi$, 
 is that $q$ satisfies the Painlev\'e-II equation and that $r(s) = q'(s)$. We denote by 
 \begin{eqnarray}
 \left(
 \begin{array}{c}
 \Phi_1(\zeta,s) \\
 \Phi_2(\zeta,s)
 \end{array} \right)
 \end{eqnarray}
 the unique solution of the Lax pair (\ref{system_lax}) which satisfies the real asymptotics
 \begin{eqnarray}\label{asympt_phi}
 \Phi_1(\zeta,s) = \cos{\left(\frac{4 \zeta^3}{3}  + s \zeta \right)} + {\cal O}(\zeta^{-1}) \;, \;
 \Phi_2(\zeta,s) = -\sin{\left(\frac{4 \zeta^3}{3}  + s \zeta \right)} + {\cal O}(\zeta^{-1}) \;,
 \end{eqnarray}
 as $\zeta \to \pm \infty$ for $s$ real. There is such a solution (see~\cite{BI03}) and it further satisfies the properties that $\Phi_1(\zeta,s)$ and
 $\Phi_2(\zeta,s)$ are real for real $\zeta$ and $s$ and 
 \begin{eqnarray}
 \Phi_1(-\zeta,s) = \Phi_1(\zeta,s) \;, \; \Phi_2(-\zeta,s) = -\Phi_2(\zeta,s) \;.
 \end{eqnarray}
One can also check from (\ref{system_lax}) that, for $s \to \infty$ and $\zeta$ real, one has
 \begin{eqnarray}\label{asympt_phi_2}
 \Phi_1(\zeta,s) \sim \cos{\left(\frac{4 \zeta^3}{3}  + s \zeta \right)} \;, \;
 \Phi_2(\zeta,s) \sim -\sin{\left(\frac{4 \zeta^3}{3}  + s \zeta \right)} \;.
 \end{eqnarray}
Note that the psi-functions initially studied by Flaschka and Newell in Ref. \cite{FN_80} are $\Phi_1 + i \Phi_2$ and $\Phi_1 - i \Phi_2$.  
From Eqs. (\ref{system_lax}, \ref{matrice_B}), one finds that $\Phi_{2}(\zeta, s)$ satisfies the interesting equation
\begin{eqnarray}\label{ev_eq_phi2}
\partial^2_s \Phi_2(\zeta,s) - (q^2-q')\Phi_2(\zeta,s) = -\zeta^2 \Phi_2(\zeta,s) \;.
\end{eqnarray}
This relation (\ref{ev_eq_phi2}), in view of the above equation (\ref{schrod_eq}) suggests to look for a solution of (\ref{eq_y_s_w}) under the form
\begin{eqnarray}\label{expr_h1_c}
f(s,w) = \int_0^\infty c(\zeta) \Phi_2(\zeta,s) e^{- w \zeta^2} \, d\zeta \;.
\end{eqnarray}
To find this function $c(\zeta)$ it is sufficient to look at the large $s$ behavior of $f(s,w=0)$. Indeed we obtained previously that
\begin{eqnarray}
f(s, w =0) \sim -\frac{8}{\pi} {\rm Ai}'(2^{-2/3}\,s) \;.
\end{eqnarray}
On the other hand from Eq. (\ref{expr_h1_c}) and Eq. (\ref{asympt_phi_2}) one has that
\begin{eqnarray}
f(s, w = 0) \sim - \int_0^\infty c(\zeta) \sin{\left(\frac{4 \zeta^3}{3}  + s \zeta \right)} d \zeta \;.
\end{eqnarray}
Now using the integral representation of the derivative of the Airy function \cite{abramowitz} 
\begin{eqnarray}
{\rm Ai}'(z) = - \frac{1}{\pi}\int_0^\infty t \sin\left(\frac{t^3}{3} + z \,t \right) \, dt \;, 
\end{eqnarray}
one recognizes that
\begin{eqnarray}
c(\zeta) = - \frac{2^{13/3}}{\pi^2} \zeta \;.
\end{eqnarray} 
One thus finds that $f(s,w)$ can be expressed in terms of $\Phi_2(\zeta,s)$ as
\begin{eqnarray}\label{explicit_h1}
f(s,w) = -\frac{2^{13/3}}{\pi^2} \int_{0}^\infty  \zeta \Phi_2(\zeta,s) e^{- w \zeta^2} \, d\zeta  \;. 
\end{eqnarray} 
We can then check, using the equations in (\ref{system_lax}) that this expression given in Eq. (\ref{explicit_h1}) is indeed a solution of the third order differential equation given in Eq. (\ref{eq_y_s_w}) and behaves asymptotically as in Eq.~(\ref{asympt_h1}). This expression (\ref{explicit_h1}) together with the Eq. (\ref{joint_explicit}) yields our main result announced in the introduction in Eqs. (\ref{explicit_P}), (\ref{explicit_h1_intro}). 

Looking at the expression (\ref{explicit_h1}) one might be worried about the convergence of this integral for $w < 0$. However, when $\zeta \to \infty$, $\Phi_2(\zeta,s)$ is a highly oscillating function, $\Phi_2(\zeta,s) \sim -\sin(\frac{4}{3} \zeta^3 + s \zeta)$, and it is then possible to give a meaning to this integral by using a standard regularization scheme, where one multiplies the integrand by $\exp{(-\epsilon|\zeta|^3)}$, with $\epsilon > 0$ and then take the limit $\epsilon \to 0^+$. One can check explicitly this procedure to obtain that the large $s$ behavior of $f(s,w)$ as defined in Eq. (\ref{explicit_h1}) yields the expected behavior in Eq. (\ref{asympt_h1}). In that case one indeed replaces $\Phi_2(\zeta,s)$ by $-\sin(\frac{4}{3} \zeta^3 + s \zeta)$, [which is its large $s$ behavior (\ref{asympt_phi_2})] and the integral over $\zeta$ in Eq. (\ref{explicit_h1}) can indeed performed, also for $w <0$, using the aforementioned regularization.

\subsection{Asymptotic analysis}
 
 Here we first give the asymptotic analysis of the joint pdf for large $s$. As we will see, this provides a way to compare our results with the results of Moreno Flores, Quastel and Remenik given in Ref. \cite{quastel_jpdf} and to obtain the amplitudes $\alpha$ and $\beta$ in Eq. (\ref{jpdf_y}). Then we obtain the asymptotic behavior of the marginal pdf $P(w)$ for large $w$.  
 
 \subsubsection{Joint probability density function for $s \to \infty$}
 
We first begin by analyzing the limit $s \to \infty$ of the jpdf ${P}(s,w)$ in Eq. (\ref{joint_explicit}). In this limit, one can obtain an estimate of the integral over $x$ in Eq. (\ref{joint_explicit}) by simply replacing $h(x,w)$ by its asymptotic behavior given in Eq. (\ref{asympt_h1}) as $x > s \gg 1$. Besides, for $s \to \infty$, one has to leading order ${\cal F}_1(s) \sim 1$. Therefore one has, for $s \to \infty$
\begin{eqnarray}\label{asympt_vicious}
{P}(s,w) \sim \int_{2^{-2/3}s}^\infty dz \left[{\rm Ai}'^2\left(\frac{w^2}{2^{8/3}} + z \right) - \frac{w^2}{2^{8/3}} {\rm Ai}^2\left(\frac{w^2}{2^{8/3}} + z \right)  \right] \;.
\end{eqnarray}

On the other hand, in Ref. \cite{quastel_jpdf}, the authors obtained the following expression for  the jpdf $\hat P(m,t)$ of the maximum (\ref{long}) and its position (\ref{trans}) of the process $Y(u)$ (\ref{def_y}) [see Eqs. (1.3), (1.4) therein]:
\begin{eqnarray}\label{expr_quastel}
\hat {P}(m,t) = 2^{1/3} {\cal F}_1(2^{2/3} m) \int_{0}^\infty dx \int_{0}^\infty dy \, \Phi_{-t, m}(2^{1/3}x) \rho_{2^{2/3}m}(x,y) \, \Phi_{t,m}(2^{1/3} y) \;,
\end{eqnarray}
where
\begin{eqnarray}\label{expr_quastel_phi}
\Phi_{t,m}(x) = 2 e^{x\, t} [t {\rm Ai}(t^2+m+x) + {\rm Ai}'(t^2+m+x)] \;, \; 
\end{eqnarray}
and 
\begin{eqnarray}
\rho_m(x,y) = (I - \Pi_0 {\cal B}_m \Pi_0)^{-1}(x,y) \;, 
\end{eqnarray}
where ${\cal B}_m$ is the integral operator defined in Eq. (\ref{kernel_ai1}) and $\Pi_0$ is the projector on $[0, \infty)$. In the large $m$ limit, one can use the estimate, $\rho_{2^{2/3}m}(x,y) \sim \delta(x-y)$ in Eq. (\ref{expr_quastel}), so that one gets the large $m$ estimate of $\hat {P}(m,t)$ as
\begin{eqnarray}\label{asympt_airy}
\hat {P}(m,t) \sim 4 \int_m^\infty  dz \left[{\rm Ai}'^2\left({t^2} + z \right) - {t^2} {\rm Ai}^2\left({t^2} + z \right)  \right] \;.
\end{eqnarray}
By comparing these two asymptotic formulas (\ref{asympt_vicious}, \ref{asympt_airy}), one finds that $\alpha$ and $\beta$ in Eq. (\ref{jpdf_y}) are given by $\alpha = 2^{2/3}$ and $\beta = 2^{4/3}$ such that
\begin{eqnarray}\label{correspondence}
\hat {P}(m,t) = 4 {P}(2^{2/3} m,2^{4/3} t) \;.
\end{eqnarray}
Interestingly, it turns out that the integral over $z$ in Eq. (\ref{asympt_vicious}) can be performed explicitly, yielding
\begin{eqnarray}
{P}(s,w) &\sim& 
\frac{1}{48} \Bigg(2^{2/3} \left(s+w^2\right) \left(4 s+w^2\right)
   \text{Ai}\left(\frac{w^2+4 s}{4\ 2^{2/3}}\right)^2- 2^{10/3} \left(s+w^2\right)
   \text{Ai}'\left(\frac{w^2+4 s}{4\ 2^{2/3}}\right)^2 \nonumber \\
 &&  -32 \text{Ai}\left(\frac{w^2+4
   s}{4\ 2^{2/3}}\right) \text{Ai}'\left(\frac{w^2+4 s}{4\ 2^{2/3}}\right)\Bigg) \;,
\end{eqnarray}
which allows to obtain straightforwardly the large $|w|$ behavior of $P(s,w)$, valid for $s \to \infty$
 \begin{eqnarray}
 \log {P}(s,w) \simeq -\frac{1}{12}|w|^3 - \frac{1}{2}|w| \, s + {\cal O}(\log |w|) \;.
 \end{eqnarray} 
 Note that the leading term matches perfectly with the behavior obtained in the large deviation regime, when $M \to \sqrt{2N}$ (\ref{cubic_largedev}), with $w \to 2^{7/3} v$. 
 
 \subsubsection{Marginal probability density function of the position of the maximum and its large $w$ behavior}
 
 We now investigate the marginal pdf $P(w)$ of the position of the maximum, which corresponds to the distribution of the end-point $X/T^{2/3}$ of the directed polymer (Fig. \ref{fig_polymer}). It is obtained from the jpdf ${P}(s,w)$ by integration over $s$:
\begin{eqnarray}\label{marginal_w}
P(w) = \frac{\pi^2}{2^{20/3}} \int_{-\infty}^\infty ds \, {\cal F}_1(s) \int_s^\infty f(x,w) f(x,-w) \, dx \;,
\end{eqnarray} 
 which is obviously symmetric $P(w) = P(-w)$, as it should be. It is normalized, thanks to the initial formula we started with (\ref{normalization}), though we could not check it directly from the above formula (\ref{marginal_w}). The distribution of the endpoint $X$ of the directed polymer of length $T$ has been widely studied in the physics literature. It was argued from scaling argument as well as from the study of simplified model, the so called "toy model" \cite{BO90,groeneboom}, that $\log P(X) \propto - (X/T^{2/3})^3$ for $X \gg T^{2/3}$. The expression above (\ref{marginal_w}) yields an exact expression for the distribution of $X/T^{2/3}$ (\ref{trans}, \ref{jpdf_y}) from which we can extract the asymptotic behavior.  
 
To analyze the large $w$ behavior of (\ref{marginal_w}) the explicit formula obtained above (\ref{explicit_h1}) is very useful. Indeed, in the limit $w \to +\infty$, the two quantities $f(x,w)$ and $f(x,-w)$ behave quite differently. For $f(x,w)$, the  integral over $\zeta$ is dominated by the region of small $\zeta$, due to the factor $e^{-w \zeta^2}$ while $f(x,-w)$ is instead dominated by the region of large $\zeta$, due to the term $e^{w \zeta^2}$. One can then obtain the leading behaviors of $f(x,w)$ and $f(x,-w)$ by estimating the small and large $\zeta$ behavior, respectively, of $\Phi_2(\zeta,x)$. The details of this analysis are left in Appendix~\ref{appendix_asymptotic}. It yields the asymptotic behavior  
\begin{eqnarray}\label{asympt_margin}
 \log P(w) = -\frac{1}{12}w^3 + o(w^3) \;. 
\end{eqnarray}
Finally, from Eq. (\ref{correspondence}) one obtains the asymptotic behavior of the pdf $\hat P(t)$ of the position of the maximum of the Airy$_2$ process minus a parabola (\ref{def_y}) as
\begin{eqnarray}
\log \hat P(t) = -\frac{4}{3}t^3 + o(t^3) \;,
\end{eqnarray}
which is in agreement with the result announced, without any detail, in Ref. \cite{quastel_jpdf}, where the amplitude of the cubic term was however not computed. Note also that this exponent $-\frac{4}{3}t^3$ was very recently given as an upper bound on the decay of the tail of $\hat P(t)$ in Ref. \cite{quastel_tail}.

\section{Conclusion}

To summarize, we have studied the joint probability density function $P_N(M,\tau_M)$ of the maximum $M$ and its position $\tau_M$ (see Fig. \ref{fig_excursions}) for $N$ non-intersecting excursions in the large $N$ limit. Once $M$ and $\tau_M$ are properly shifted and rescaled we have shown that the jpdf converges, when $N \to \infty$, to a jpdf $P(s,w)$ (\ref{convergence}) which we have computed explicitly (\ref{joint_explicit}). This jpdf $P(s,w)$ yields, up to a rescaling (\ref{correspondence}), the jpdf of the maximum and its position of the Airy$_2$ process minus a parabola~(\ref{def_y}). The expression obtained here is actually different from the one (\ref{expr_quastel}) recently obtained in Ref. \cite{quastel_jpdf}. In particular we show here that this jpdf can be expressed in terms of the psi-function for the Hastings-McLeod solution to the Painlev\'e II equation. After this paper has been submitted, the equivalence of the two formulas was shown in Ref. \cite{BLS12}.

From $P(s,w)$ one obtains an exact expression for the marginal pdf $P(w)$, from which we have obtained the asymptotic behavior $\log P(w) \sim -w^3/12$, establishing on firmer grounds this cubic behavior which had been proposed some time ago in the physics literature, based on scaling argument \cite{halpin_review}, the analysis of the "toy" model \cite{BO90} and numerical simulations \cite{halpin_pra,goldschmidt}. Very recently, this pdf was also measured experimentally, indicating also such a cubic behavior in the tail of the pdf, while a more precise comparison between our exact result and the experimental data would certainly be very interesting. 

\begin{acknowledgement}
It is a pleasure to thank A. Comtet, P. J. Forrester, S. N. Majumdar and J. Rambeau for our fruitful collaborations on this subject and K. Takeuchi for sharing his experimental data. I would also like to acknowledge K. Johansson, J. Krug for stimulating discussions, T. Claeys, A. Its and in particular K. Liechty for very useful correspondence. This research was 
supported by ANR grant 2011-BS04-013-01 WALKMAT and in part by the Indo-French 
Centre for the Promotion of Advanced Research under Project 4604-3. 
\end{acknowledgement}

\appendix

\section{Asymptotic behavior of the marginal probability density function $P(w)$}\label{appendix_asymptotic}

In this appendix, we study the asymptotic behavior of the (marginal) pdf $P(w)$ of the position of the maximum. The starting point of our analysis is the formula given in the text in Eq.~(\ref{marginal_w}), which we recall here
\begin{eqnarray}\label{marginal_w_app}
P(w) = \frac{\pi^2}{2^{20/3}} \int_{-\infty}^\infty ds \, {\cal F}_1(s) \int_s^\infty f(x,w) f(x,-w) \, dx \;.
\end{eqnarray}
We first obtain the leading behaviors of $f(x,w)$ and $f(x,-w)$ (\ref{explicit_h1}) by estimating the small and large $\zeta$ behavior, respectively, of $\Phi_2(\zeta,x)$. 

One can obtain the small $\zeta$ behavior of $\Phi_2(\zeta,x)$ by analyzing the coupled equations for $\Phi_1(\zeta,x), \Phi_2(\zeta,x)$ in (\ref{system_lax}, \ref{matrice_A}) and using that $\Phi_1(-\zeta,x) = \Phi_1(\zeta,x)$ while $\Phi_2(-\zeta,x) = -\Phi_2(\zeta,x)$. One obtains
\begin{eqnarray}
&&\Phi_1(\zeta,x) = C e^{-\int_x^\infty q(y) dy} + {\cal O}(\zeta^2) \;, \\
&&\Phi_2(\zeta,x) = -\zeta e^{\int_x^\infty q(y) dy} [C \int_{-\infty}^x dz \, e^{-2 \int_z^\infty q(y) dy} - D] + {\cal O}(\zeta^3)\;,
\end{eqnarray} 
where at this stage the constants $C$ and $D$ remains undetermined. The large $w > 0$ behavior of $f(x,w)$ is then given by
\begin{eqnarray}\label{h1_largew+}
f(x,w) = -\frac{2^{13/3}}{\pi^2} \int_0^\infty \zeta \phi^2(\zeta,x) e^{-w \zeta^2} d \zeta \sim \frac{2^{13/3}}{4 \pi^2 w^{3/2}} e^{\int_x^\infty dy q(y) dy} [C \int_{-\infty}^x dz \, e^{-2 \int_z^\infty q(y) dy} - D] \;.
\end{eqnarray}
The constants $C$ and $D$ can then be obtained by extracting the large $x$ behavior of the above expression (\ref{h1_largew+}) and matching it with the large $w$ expansion of the expression given in Eq. (\ref{asympt_h1}), assuming that the limits $w \to \infty$ and $x \to \infty$ do commute. When $x\to \infty$ the integral over $z$ in Eq. (\ref{h1_largew+}) is diverging and it is easy to see that one has
\begin{eqnarray}\label{inter}
C \int_{-\infty}^x dz \, e^{-2 \int_z^\infty q(y) dy}  \sim C x \;, \; x \gg 1 \;.
\end{eqnarray}
Therefore, to leading order in $x$, for $x \to \infty$, one obtains from Eqs. (\ref{h1_largew+}) and (\ref{inter})
\begin{eqnarray}
f(x,w) \sim \frac{2^{7/3} \,C \, x}{\pi^2 w^{3/2}} \;, 
\end{eqnarray}
while the expression in Eq. (\ref{asympt_h1}) yields
\begin{eqnarray}\label{asympt_h1_app}
f(x,w) = \frac{2^{7/3} x}{\pi^{3/2} w^{3/2}} + {\cal O}(w^{-5/2})\;,
\end{eqnarray}
which, by comparison, yields immediately
\begin{eqnarray}
C = \sqrt{\pi} \;.
\end{eqnarray}
The computation of $D$ is slightly more involved but by comparing Eq. (\ref{h1_largew+}) in the large $x$ limit one obtains
\begin{eqnarray}\label{expr_C}
D = C \left[ \int_{-\infty}^0 dz e^{-2 \int_z^\infty q(y) dy} +  \int_0^\infty dz (1-e^{-2\int_z^\infty q(y) dy}) \right]\;.
\end{eqnarray} 
Given that $q(y) > 0$, for all $y$ real, one sees immediately on that expression (\ref{expr_C}) that $D >0$.

To compute the asymptotic behavior of $f(x,-w)$ for $w \to \infty$ we notice that, in this case, the behavior of the integral over $\zeta$ in Eq. (\ref{explicit_h1}) is instead dominated by the region $\zeta \to \infty$ where, to leading order in $w$, $\Phi_2(\zeta,x)$ can thus be replaced by its asymptotic behavior given in Eq. (\ref{asympt_phi_2}). Performing the integral over $\zeta$ one thus arrives at the expression obtained previously in Eq. (\ref{asympt_h1}), with the substitution $w \to - w$ from which one has to extract carefully the large $w$ behavior. One obtains
\begin{eqnarray}\label{h1_largew-}
f(x,-w) \sim -\frac{2^{7/3}}{\pi^{3/2}} w^{1/2} e^{-\frac{w^3}{12} - \frac{wx}{2}} \;.
\end{eqnarray}
These asymptotic behaviors in Eqs. (\ref{h1_largew+}, \ref{h1_largew-}) suggest to separate the integral over $s$ in Eq. (\ref{marginal_w_app}) into two parts, 
\begin{eqnarray}
&&P(w) = P_+(w) + P_-(w) \nonumber \\
&&P_+(w) = \frac{\pi^2}{2^{20/3}} \int_{0}^\infty ds \, {\cal F}_1(s) \int_s^\infty f(x,w) f(x,-w) \, dx \;, \label{def_p+} \\
&&P_-(w) = \frac{\pi^2}{2^{20/3}} \int_{-\infty}^{0} ds \, {\cal F}_1(s) \int_s^\infty f(x,w) f(x,-w) \, dx  \label{def_p-} \;. \\
\end{eqnarray}
Let us first analyze $P_+(w)$ in Eq. (\ref{def_p+}). There, because of the exponential term $e^{-w \,x/2}$ coming from (\ref{h1_largew-}) one can perform, 
in the integral over $x$, the change of variable $z = w x$ to obtain
\begin{eqnarray}\label{inter_p+}
\int_s^\infty f(x,w) f(x,-w) \, dx \sim \frac{\tilde c}{w^{2}} e^{-\frac{1}{12}w^3 - \frac{ws}{2}} \;,
\end{eqnarray} 
where $\tilde c$ can be read from Eqs. (\ref{h1_largew+}, \ref{h1_largew-}):
\begin{eqnarray}
\tilde c= \frac{2^{9/2}}{\pi^3} \int_{0}^\infty dz \, (1-e^{-2 \int_z^\infty q(y) dy}) \;,
\end{eqnarray}
where we have used the explicit expression of $C$ in Eq. (\ref{expr_C}). 
 Finally, from Eq. (\ref{inter_p+}) one obtains the large $w$ behavior of $P_+(w)$, to leading order as
 \begin{eqnarray}\label{final_p+}
 P_+(w) \sim \frac{\tilde C}{w^3} {e^{-\frac{1}{12}w^3}} \;, \; \tilde C = \frac{2^{-7/6}}{\pi} {\cal F}_1(0) \int_0^{+\infty} dz \, (1-e^{-2 \int_z^\infty q(y) dy}) \;.
 \end{eqnarray}
 
 Let us now analyse $P_-(w)$ in Eq. (\ref{def_p-}), whose analysis turns out to be more complicated. We first notice that ${\cal F}_1(s)$ is bounded for $s < 0$ and in addition, its asymptotic behavior for $s \to -\infty$ is given by \cite{BBD08,borot_satya}
\begin{eqnarray}
{\cal F}_1(s) = \frac{2^{-11/48}}{|s|^{1/16}} e^{\zeta'(-1)/2}e^{\frac{-|s|^3}{24}-\frac{|s|^{3/2}}{3 \sqrt{2}}}(1 + {\cal O}(|s|^{-3/2})) \;,
\end{eqnarray}
so that there exists a constant $K$ such that 
\begin{eqnarray}
{\cal F}_1(s) \leq K e^{-s^2} \;, \; \forall s < 0 \;.
\end{eqnarray}
 Therefore one has
 \begin{eqnarray}\label{def_i-}
 P_-(w) \leq I_-(w) \;, \; I_-(w) = K \int_{-\infty}^0 ds \, e^{-s^2} \int_s^\infty f(x,w) f(x,-w) \, dx \;.
 \end{eqnarray}
 For large $w$, because of the exponential term $e^{-w x/2}$ in Eq. (\ref{h1_largew-}), the integral over $s$ in Eq. (\ref{def_i-}) is dominated by the region of large negative $s$. On the other hand, for large negative $s$, the integral over $x$ is also dominated by the region where $x$ is large and negative. Now using the behavior of $q(s)$ (see {\it e.g.} Ref. \cite{BBD08})
\begin{eqnarray}
q(s) \sim \sqrt{\frac{-s}{2}} \;, \; s \to -\infty \;,
\end{eqnarray}  
 one obtains that, for large negative $s$, one has
 \begin{eqnarray}\label{estimate}
 \int_s^\infty f(x,w) f(x,-w) \, dx \sim \frac{d}{w}e^{-\frac{1}{12}w^3 + \frac{\sqrt{2}|s|^{3/2}}{3}- \frac{ws}{2}}  \;,
 \end{eqnarray}
 where $d > 0$ is a constant which can be read from Eqs. (\ref{h1_largew+}, \ref{h1_largew-}). Using this last estimate (\ref{estimate}), one immediately sees that the large $w$ behavior $I_-(w)$ in Eq. (\ref{def_i-}) is given by
 \begin{eqnarray}\label{final_i-}
 \log I_-(w) = -\frac{1}{12} w^3 + o(w^3) \;.
 \end{eqnarray}
 Finally, using the estimates in Eq. (\ref{final_p+}) and (\ref{final_i-})  together with the inequality in Eq. (\ref{def_i-}) one obtains
 \begin{eqnarray}
 \log P(w) = -\frac{1}{12} w^3 + o(w^3) \;,
 \end{eqnarray}
 as given in the text (\ref{asympt_margin}).


\begin{thebibliography}{}

\bibitem{halpin_review}
T. Halpin-Healy, Y.C. Zhang, {\it Kinetic roughening phenomena, stochastic growth, directed polymers and all that}, Phys. Rep. {\bf 254}, 215 (1995).

\bibitem{huse_henley}
D.~A. Huse, C.~L. Henley, {\it Pinning and roughening of domain walls in Ising systems due to random impurities}, Phys. Rev. Lett. {\bf 54}, 2708 (1985).

\bibitem{kardar_dprm}
M. Kardar, {\it Depinning by Quenched Randomness}, Phys. Rev. Lett. {\bf 55}, 2235 (1985).

\bibitem{kpz}
M. Kardar, G. Parisi, Y.C. Zhang, {\it Dynamic Scaling of Growing Interfaces}, Phys. Rev. Lett. {\bf 56}, 889 (1986).

\bibitem{burgers}
D. Forster, D.~R. Nelson, M.~J. Stephen, {\it Large-distance and long-time properties of a randomly stirred fluid}, Phys. Rev. A {\bf 16}, 732 (1977).

\bibitem{mezard_dprm}
M. M{\'e}zard, {\it On the glassy nature of random directed polymers in two dimensions}, J. Phys-Paris {\bf 51}, 1831 (1990).

\bibitem{hwa_lassig}
T. Hwa, M. L{\"a}ssig, {\it Similarity Detection and Localization}, Phys. Rev. Lett. {\bf 76}, 2591 (1996).
\bibitem{lemerle}
S. Lemerle, J. Ferr\'e, C. Chappert, V. Mathet, T. Giamarchi, P. Le Doussal, {\it Domain Wall Creep in an Ising Ultrathin Magnetic Film}, Phys. Rev. Lett. {\bf 80}, 849 (1998). 


\bibitem{moulinet}
S. Moulinet, A. Rosso, W. Krauth, E. Rolley, {\it Width distribution of contact lines on a disordered substrate}, Phys. Rev. E {\bf 69}, 035103(R) (2004).

\bibitem{johansson_dprm}
K. Johansson, {\it Discrete polynuclear growth and determinantal processes}, Comm. Math. Phys. {\bf 242}, 277 (2003).

\bibitem{TW96}
C.~A. Tracy, H.~Widom,  {\it On orthogonal and symplectic matrix ensembles}, Commun. Math. Phys. \textbf{177}, 727 (1996).




\bibitem{krug_pra}
J. Krug, P. Meakin, T. Halpin-Healy, {\it Amplitude universality for driven interfaces and directed polymers in random media}, Phys. Rev. A {\bf 45}, 638 (1992).  

\bibitem{baik_rains}
J. Baik, E. Rains, {\it Symmetrized random permutations}, in {\it Random Matrix Models and Their Applications}, edited by P. Bleher and A. Its, MSRI Publications 40, Cambridge University Press, 2001.

\bibitem{spohn_praehofer_prl}
M. Pr\"ahofer, H. Spohn, {\it Universal Distributions for Growth Processes in 1+1 Dimensions and Random Matrices}, Phys. Rev. Lett. {\bf 84}, 4882 (2000). 


\bibitem{forrester_npb}
P. J. Forrester, S. N. Majumdar, G. Schehr, {\it Non-intersecting Brownian walkers and Yang-Mills theory on the sphere}, Nucl. Phys. B {\bf 844}, 500 (2011); Erratum Nucl. Phys. B {\bf 857}, 424 (2011). 


\bibitem{liechty}
K. Liechty, {\it Nonintersecting Brownian motions on the half-line and discrete Gaussian orthogonal polynomials}, J. Stat. Phys. {\bf 147}, 582 (2012)

\bibitem{quastel_jpdf}
G. R. Moreno Flores, J. Quastel, D. Remenik, {\it Endpoint distribution of directed polymers in 1+1 dimensions}, preprint arXiv:1106.2716.  

\bibitem{johansson_2000}
K. Johansson, {\it Shape Fluctuations and Random Matrices}, Commun. Math. Phys. {\bf 209}, 437 (2000). 





\bibitem{spohn_praehofer}
M. Pr\"ahofer, H. Spohn, {\it Scale Invariance of the PNG Droplet and the Airy Process}, J. Stat. Phys., {\bf 108}, 1071 (2002).  


\bibitem{TW94a}
C.~A. Tracy, H.~Widom, \emph{Level-spacing distributions and the {Airy}
  kernel}, Commun. Math. Phys. \textbf{159} (1994), 151--174.


\bibitem{BDJ}
J. Baik, P. Deift, K. Johansson, {\it On the distribution of the length of the
longest increasing subsequence of random permutations}, J. Amer. Math. Soc. {\bf 12}, 1119 (1999).


\bibitem{satya_review}
S.~N. Majumdar, {\it Random matrices, the Ulam problem, directed polymers and
growth models, and sequence matching in Complex Systems}, (Les Houches lecture
notes ed. by J.-P. Bouchaud, M. M{\'e}zard, and J. Dalibard) (Elsevier,
Amsterdam), 179 (2007).


\bibitem{calabrese_epl}
P. Calabrese, P. Le Doussal, A. Rosso, {\it Free-energy distribution of the directed polymer at high temperature}, Europhys. Lett. {\bf 90}, 20002 (2010).


\bibitem{dotsenko_epl}
V. Dotsenko, {\it Bethe ansatz derivation of the Tracy-Widom distribution for one-dimensional directed polymers}, Eutrophys. Lett. {\bf 90}, 20003 (2010).  




\bibitem{dotsenko_jstat}
V. Dotsenko, {\it Replica Bethe ansatz derivation of the Tracy-Widom distribution of the free energy fluctuations in one-dimensional directed polymers}, J. Stat. Mech. P07010 (2010). 

\bibitem{sasamoto_spohn1}
T. Sasamoto, H. Spohn, {\it The one-dimensional KPZ equation: an exact solution and its universality}, Phys. Rev. Lett. {\bf 104}, 230602 (2010).


\bibitem{sasamoto_spohn2}
T. Sasamoto, H. Spohn, {\it Exact height distributions for the KPZ equation with narrow wedge initial condition}, Nucl. Phys. B {\bf 834}, 523 (2010). 


\bibitem{sasamoto_spohn3}
T. Sasamoto, H. Spohn, {\it The 1+1-dimensional Kardar-Parisi-Zhang equation and its universality class}, J. Stat. Phys. {\bf 140}, 209 (2010).


\bibitem{amir_cmp}
G. Amir, I. Corwin, J. Quastel, {\it Probability Distribution of the Free Energy of the Continuum Directed Random Polymer in 1+1 dimensions}, Comm. Pure Appl. Math {\bf 64}, 466 (2011). 


\bibitem{prolhac_spohn1}
S. Prolhac, H. Spohn, {\it Two-point generating function of the free energy for a directed polymer in a random medium} J. Stat. Mech., P01031 (2011).


\bibitem{prolhac_spohn2}
S. Prolhac, H. Spohn, {\it The One-dimensional KPZ Equation and the Airy Process} J. Stat. Mech. (2011) P03020;



\bibitem{calabrese_flat}
P. Calabrese, P. Le Doussal, {\it An exact solution for the KPZ equation with flat initial conditions}, Phys. Rev. Lett. {\bf 106}, 250603 (2011). 

\bibitem{calabrese_flat_long}
P. Le Doussal, P. Calabrese, {\it The KPZ equation with flat initial condition and the directed polymer with one free end}, J. Stat. Mech., P06001 (2012).


 



\bibitem{ferrari_spohn}
P. L. Ferrari, H. Spohn, {\it A determinantal formula for the GOE Tracy-Widom distribution}, J. Phys. A: Math. Gen. {\bf 38}, L557 (2005). 


\bibitem{tracy_excursions}
C. A. Tracy, H. Widom, {\it Nonintersecting Brownian excursions}, Ann. of App. Proba., {\bf 17} (3), 953 (2007).

\bibitem{corwin_hammond}
I. Corwin, A. Hammond, {\it Brownian Gibbs property for Airy line ensembles}, arXiv:1108.2291.





\bibitem{ferrari_begrohu}
P. Ferrari, Lecture Notes of Beg-Rohu Summer School, available at http://ipht.cea.fr/Meetings/BegRohu2008/.

\bibitem{FN_80}
H. Flaschka, A.~C. Newell, {\it Monodromy and spectrum-preserving deformations I}, Comm. Math. Phys. {\bf 76}(1), 65 (1980).


\bibitem{BI03}
P. Bleher, A. Its, {\it Double scaling limit in the random matrix model: the Riemann-Hilbert approach}, Comm. Pure Appl. Math. {\bf 56}, 433 (2003).


\bibitem{abramowitz}
M. Abramowitz, I. Stegun, Editors, {\it Handbook of Mathematical Functions}, National Bureau of
Standards, Washington D.C., 10th Printing, (1972).

\bibitem{schehr_prl}
G. Schehr, S.~N. Majumdar, A. Comtet, J. Randon-Furling, {\it Exact distribution of the maximal height of p vicious walkers}, Phys. Rev. Lett. {\bf 101}, 150601 (2008). 


\bibitem{rambeau_epl}
J.~Rambeau, G.~Schehr, {\it Extremal statistics of curved growing interfaces in 1+1 dimensions},  Europhys. Lett. {\bf 91}, 60006 (2010).





\bibitem{rambeau_pre}
J.~Rambeau, G.~Schehr, {\it Distribution of the time at which N vicious walkers reach their maximal height}, Phys. Rev. E {\bf 83}, 061146 (2011). 


\bibitem{katori_jstatphys}
M. Katori, M. Izumi, N. Kobayashi, {\it Two Bessel Bridges Conditioned Never to Collide, Double Dirichlet Series, and Jacobi Theta Function}, J. Stat. Phys. {\bf 131}, 1067 (2008). 



\bibitem{kobayashi_pre}
N. Kobayashi, M. Izumi, M. Katori, {\it Maximum distributions of bridges of noncolliding Brownian paths}, Phys. Rev. E {\bf 78}, 051102 (2008). 

\bibitem{feierl}
T. Feierl, {\it The height of watermelons with wall}, J. Phys. A: Math. Theor. {\bf 45}, 095003 (2012).




\bibitem{GM94}
D. J. Gross, A. Matytsin, {\it Instanton Induced Large $N$ Phase Transitions in Two and Four Dimensional QCD}, Nucl. Phys. B. \textbf{429}, 50 (1994).

\bibitem{CNS96}
M. Crescimanno, S.G. Naculich, H.~J. Schnitzer, {\it Evaluation of the free
energy of two-dimensional Yang-Mills theory}, Phys. Rev. D {\bf 54}, 1809 (1996).





\bibitem{deharo}
S. de Haro, M. Tierz, {\it Brownian motion, Chern-Simons theory, and 2d Yang-Mills}, Phys. Lett. B {\bf 201}, 201 (2004).




  
\bibitem{GW80}  
D.~J. Gross, E.~Witten, {\it Possible third-order phase transition in the large-n
lattice gauge limit}, Phys. Rev. D {\bf 21}, 446 (1980).

\bibitem{W80}
S.~R. Wadia, {\it N = 1 phase transition in a class of exactly soluble model lattice gauge theories}, Phys. Lett. {\bf 93}B, 403 (1980).  


\bibitem{DK93}
M.~R. Douglas, V.~A. Kazakov, \emph{Large $N$ phase transition in continuum
QCD${}_2$}, Phys. Lett. B 
  \textbf{319}, 219 (1993).


\bibitem{PS90a}
V.~Periwal, D.~Shevitz, \emph{Unitary-matrix models as exactly solvable
  string theories}, Phys. Rev. Lett. \textbf{64}, 1326 (1990).



\bibitem{BO90}
J.~-P. Bouchaud, H. Orland, {\it On the Bethe ansatz for random directed polymers}, J. Stat. Phys. {\bf 61}, 877 (1990). 


\bibitem{pld_monthus}
P. Le Doussal, C. Monthus, {\it Exact solutions for the statistics of extrema of some random 1D landscapes, Application to the equilibrium and the dynamics of the toy model}, Physica A {\bf 317}, 140 (2003). 

\bibitem{groeneboom}
P. Groeneboom, (1989) {\it Brownian motion with a parabolic drift and Airy functions.}, Probab Theory Related Fields {\bf 81}, 31 (1989). 



\bibitem{halpin_pra}
T. Halpin-Healy, {\it Directed polymers in random media: Probability distributions}, Phys. Rev. A {\bf 44}, R3415 (1991).


\bibitem{goldschmidt}
Y. Y. Goldschmidt, T. Blum, {\it Directed polymers in a random medium: Universal scaling behavior of the probability distribution}, Phys. Rev. E {\bf 47}, R2979 (1993).



\bibitem{krug_review}
T. Kriecherbauer, J. Krug, {\it A pedestrian's view on interacting particle systems, KPZ universality, and random matrices}, J. Phys. A: Math. Theor. {\bf 43}, 403001 (2010). 



\bibitem{kazz_prl}
K. A. Takeuchi, M. Sano, {\it Universal Fluctuations of Growing Interfaces: Evidence in Turbulent Liquid Crystals}, Phys. Rev. Lett. 104, 230601 (2010).  
  
\bibitem{kazz_nature}  
K. A. Takeuchi, M. Sano, T. Sasamoto, H. Spohn, {\it Growing interfaces uncover universal fluctuations behind scale invariance}, Sci. Rep. (Nature) {\bf 1}, 34 (2011).  

\bibitem{kazz_evs}
K. A. Takeuchi, M. Sano, {\it Evidence for geometry-dependent universal fluctuations of the Kardar-Parisi-Zhang interfaces in liquid-crystal turbulence}, 
J. Stat. Phys. {\bf 147}, 853 (2012). 

\bibitem{degennes}
P.~-G. de Gennes, {\it Soluble Models for fibrous structures with steric constraints}, J. Chem. Phys. {\bf 48}, 2257 (1968). 





\bibitem{nadal}
C. Nadal, S. N. Majumdar, {\it A simple derivation of the Tracy-Widom distribution of the maximal eigenvalue of a Gaussian unitary random matrix}, J. Stat. Mech., P04001 (2011). 


\bibitem{szego}
G. Szeg{\"o}, {\it Orthogonal polynomials}, American Mathematical Society, Providence R.I., 4th edition, (1975).

















\bibitem{plancherel}
M. Plancherel, W. Rotach, {\it Sur les valeurs asymptotiques des polyn\^omes d'Hermite}, Comm. Math. Helv. {\bf 1}, 227 (1929).



\bibitem{mehta_book}
M. L. Mehta, {\it Random matrices}, Academic Press, New York, (1991).


\bibitem{forrester_book}
P.~J. Forrester, \emph{Log-gases and random matrices}, Princeton University
  Press, Princeton, NJ, 2010.








\bibitem{eynard_harnad}
M. Bertola, B. Eynard, J. Harnad, {\it Partition functions for Matrix Models and Isomonodromic Tau functions}, J.~Phys.~A {\bf 36}, 3067 (2003). 




\bibitem{satya_private}
S. N. Majumdar, private communication. 




\bibitem{praehofer_spohn}
M. Pr\"ahofer, H. Spohn, {\it Exact scaling functions for one-dimensional stationary KPZ growth}, J. Stat. Phys. {\bf 115} (1-2), 255 (2004).


\bibitem{forrester_spike}
P. J. Forrester, {\it Probability densities and distributions for spiked Wishart $\beta$-ensembles}, arXiv:1101.2261.


\bibitem{claeys_thesis}
T. Claeys, PhD-Thesis, "Universality in critical random matrix ensembles and pole-free solutions of Painlev\'e equations", Katholieke Universiteit Leuven (2006). 



  
  
\bibitem{quastel_tail}
J. Quastel, D. Remenik, {\it Tails of the endpoint distribution of directed polymers}, http://arxiv.org/abs/1203.2907.


\bibitem{BLS12}
J. Baik, K. Liechty, G. Schehr, {\it On the joint distribution of the maximum and its position of the Airy2 process minus a parabola}, preprint arXiv:1205.3665.


\bibitem{BBD08}
J. Baik, R. Buckingham, J. DiFranco, {\it Asymptotics of Tracy-Widom distributions
and the total integral of a Painlevé II function}, Comm. Math. Phys. {\bf 280}(2), 463 (2008). 

\bibitem{borot_satya}
G. Borot, B. Eynard, S. N. Majumdar, C. Nadal, {\it Large deviations of the maximal eigenvalue of random matrices}, J. Stat. Mech., P11024 (2011).  









\end{thebibliography}
\end{document}